\documentclass{iopart}

\usepackage{graphicx}

\jl{6}        % submitted to journal: Classical and Quantum Gravity
\eqnobysec    % section equation numbering

\def\beq{\begin{equation}}
\def\eeq{\end{equation}}
\def\eqref#1{(\ref{#1})}
\begin{document}

\title[Cylindrical gravitational waves]
{Cylindrical gravitational waves: C-energy, super-energy and associated dynamical effects}

\author{
Donato Bini${}^{\dag}{}^{\ddag}$,
Andrea Geralico${}^{\dag}$,
Wolfango Plastino${}^{*}{}^{\ddag}$
}
\address{
  ${}^{\dag}$\
 Istituto per le Applicazioni del Calcolo \lq\lq M. Picone," I-00185 Rome, Italy
}
\address{
 ${}^{\ddag}$\
 INFN, Sezione di Roma Tre, I-00146 Rome, Italy
}
\address{
 ${}^{*}$\
 Roma Tre University, Department of Mathematics and Physics, I-00146 Rome, Italy\\ 
}

\ead{donato.bini@gmail.com}

\begin{abstract}
The energy content of cylindrical gravitational wave spacetimes is analyzed by considering two local descriptions of energy associated with the gravitational field, namely those based on the C-energy and the Bel-Robinson super-energy tensor.
A Poynting-Robertson-like effect on the motion of massive test particles, beyond the geodesic approximation, is discussed, allowing them to interact with the background field through an external force which accounts for the exchange of energy and momentum between particles and waves.
In addition, the relative strains exerted on a bunch of particles displaced orthogonally to the direction of propagation of the wave are examined, providing invariant information on spacetime curvature effects caused by the passage of the wave. 
The explicit examples of monochromatic waves with either a single or two polarization states as well as pulses of gravitational radiation are discussed.
\end{abstract}

\pacno{04.20.Cv}

\section{Introduction}

Cylindrical gravitational waves have been studied since the very beginning of General Relativity by A. Einstein himself in a famous paper written in collaboration with N. Rosen in 1937 \cite{Einstein:1937qu}. The Einstein-Rosen waves constitute a simple example of a propagating gravitational field with cylindrical symmetry for which an exact solution exists. Other solutions have been investigated since then for both vacuum and electro-vacuum spacetimes as well as in the presence of perfect fluid sources, either in the form of monochromatic (elementary) waves or as wave packets \cite{Stephani:2003tm}.
Numerical studies of general solutions also exist \cite{Piran:1985dk,Celestino:2015hla,Mishima:2017zpn}.
The asymptotic behavior of cylindrically symmetric spacetimes including Einstein-Rosen waves has been deeply discussed in Refs. \cite{Stachel:1966,Ashtekar:1996cd,Ashtekar:1996cm} (see also references therein for a complete list of related works).

Einstein-Rosen waves are characterized by a single polarization and propagate outwards from a singularity located on the symmetry axis, which can be interpreted as being a material source.
These waves have been shown in Refs. \cite{Boardman:1959,Rosen:1993ea} to carry off energy and momentum by using the canonical energy-momentum pseudotensor.
The emission of gravitational radiation at spatial infinity in more general cylindrical gravitational wave spacetimes with two polarization modes has been studied in Ref. \cite{Stachel:1966} through the method of news functions associated with each degree of freedom of the waves. The cylindrical analog of the Bondi mass was also defined there in terms of the asymptotic value of the C-energy introduced by Thorne \cite{thorne_65} as the total gravitational energy per unit length along the symmetry axis enclosed by a cylinder of a given radius at a given time. Thorne's interpretation was later supported by Chandrasekhar \cite{chandra:1986}, who defined a Hamiltonian density from the Lagrangian density leading to the Ernst equation. 
However, the C-energy differs from the definition of quasi-local energy obtained by Hamiltonian methods \cite{Ashtekar:1993ds,Brown:2000dz,Goncalves:2002tx}, the agreement being recovered in the weak field limit only.

Thorne also defined a C-energy flux four vector, which is conserved and has associated energy density and spatial momentum with respect to a given family of observers. 
Such quantities naturally enter the canonical quantization of Einstein-Rosen waves as canonically conjugate variables to time and radius \cite{Kuchar:1971xm}.  
Regardless of what is the correct definition of energy density, the C-energy can be considered as a powerful tool in the analysis of both local and global dynamics of systems endowed with cylindrical symmetry.
Another useful mathematical tool largely used to describe the energy content of a given spacetime is represented by the Bel and Bel-Robinson super-energy tensors \cite{bel58,bel59,bel62,rob59,rob97}.
The latter were introduced for vacuum gravitational fields in order to provide a definition of local energy density (super-energy density) and energy flux (super-Poynting vector) in complete formal analogy with electromagnetism.
Their properties have been extensively investigated over the years and applications to many different contexts, including gravitational wave spacetimes, have been discussed \cite{Deser:1977zz,Breton:1993du,Mashhoon:1996wa,Bonilla:1997,Bergqvist:1998b,Mashhoon:1998tt,Senovilla:1999xz,bini-jan-miniutti,Lazkoz:2003wx,Herrera:2006vu,GomezLobo:2007gm,Ferrando:2008sr,Gomez-Lobo:2013eya,Bini:2018lhh,Dominguez:2018evg}. 
A conserved (super-momentum) four vector can be formed also in this case \cite{Maartens:1997fg,Senovilla:1999xz}, carrying the twofold information of the super-energy density and the spatial super-momentum density as measured by a given observer family. 
Both C-energy and super-energy densities thus have an observer-dependent meaning. 

Here we are interested in studying the energy content of cylindrical gravitational waves, by comparing the super-energy tensor description of the local properties of the gravitational field with that related to the C-energy. 
These two definitions of energy are indeed substantially different and do not share the same geometrical meaning.
In fact, the super-energy tensors are defined in terms of the spacetime curvature, whereas the C-energy is formulated in terms of metric quantities, but has also a true geometrical (gauge-invariant) meaning being related to the area element spanned by the two commuting Killing vectors associated with the cylindrical symmetry, $\partial_z$ (with $z$ the symmetry axis) and $\partial_\phi$ (generating rotation around it).
However, since the metric tensor satisfies a wave equation, the second derivatives of the metric (super-energy) should be (roughly) proportional to the metric itself (C-energy), so that one could expect that both approaches bear a similar information, at least qualitatively. 
An analytic and systematic comparison in this sense has never been performed.

To this end, we will investigate a dynamical effect allowing massive test particles to interact with the background gravitational field {\it \`a la} Poynting-Robertson \cite{Poy,Rob}. These authors first investigated the effect of the radiation pressure of the light emitted by a star on the motion of small bodies orbiting it, by assuming a Thomson-type interaction, i.e., absorption and consequent re-emission of radiation by the particle, leading to a non-geodesic motion.
Following such an approach, the scattering of test particles by radiation fields was studied in several backgrounds of astrophysical interest \cite{abram,ML,Bini:2008vk,Oh:2010qn,Bini:2011zza,Bini:2014sua,Bini:2011zz,Bini:2014yca}, and variations of it were also considered to account for different kinds of interaction \cite{Bini:2012ff,Bini:2012zzd,Bini:2013es,Bini:2014lwz}.
We will use here the same mathematical prescription, mimicking the interaction with the background field by means of an external force entering the equations of motion,  associated with the energy and momentum fluxes carried by the waves. 
This force is defined here in terms of the 4-momentum density of radiation observed in the particle's rest frame by a (constant) effective interaction cross section.

We will consider three explicit solutions belonging to this class of spacetimes, for which the underlying mathematical analysis results much simplified and allows for comparison: Einstein-Rosen waves \cite{Einstein:1937qu}, Chandrasekhar's solution for cylindrical monochromatic waves with two
polarizations \cite{chandra:1986}, and gravitational wave pulses (Weber-Wheeler-Bonnor solution \cite{Weber:1957oib,Bonnor:1957}).
We will first recall the main geometrical properties of such spacetimes, with special attention to Chandrasekhar's solution, which seems to be poorly studied in the literature.
We will then study the geodesic motion of massive particles as well as the world line deviation between neighboring particles as measured by observers at rest with respect to the coordinates chosen as \lq\lq fiducial'' observers. We will use the definition of relative accelerations and strains among a set of comoving particles discussed, e.g., in Ref. \cite{Bini:2006vp,Bini:2007zzb} to study the deformation of a ring of particles initially at rest in a plane orthogonal to the direction of propagation of the wave.
Finally, we will investigate the features of the Poynting-Robertson-like effect described above by comparing the C-energy and super-energy fluxes as representative of the momentum exchange between particles and background field, and hence of the associated acceleration effects, within this approximation.
This analysis would help to better characterize these two energy notions not only from a mathematical (symmetry and geometrically related) point of view, but more interestingly from a physical point of view, accounting for the dynamical effects associated with them.

The paper is organized as follows. In Section II we shortly review the description of cylindrical gravitational wave spacetimes and introduce a family of fiducial observers and adapted frames.
In Section III we write down the geodesic equations for massive particles, with particular attention to radial geodesics, which are the only ones allowed to exist in a plane orthogonal to the symmetry axis in the most general case. 
The definition of relative acceleration and strains among a set of comoving test particles is given in Section IV.
In Section V we then provide the definition of C-energy and super-energy densities, used to discuss the acceleration effects associated with the energy-momentum exchange between particles and background field of the wave in Section VI.
In Section VII we apply the above results to some explicit solutions belonging to this class of spacetimes, describing monochromatic waves as well as gravitational wave pulses. 
Finally, Section VIII summarizes the outcome of the present work.

We use the notation and conventions of Ref. \cite{Misner:1974qy}, with the metric signature mostly positive ($-+++$), Greek indices running from 0 to 3 and Latin ones from 1 to 3. Moreover, we use natural units such that $c=G=1$.

\section{Cylindrical gravitational waves}

In the literature there are many variations for the cylindrically symmetric line element in cylindrical-like coordinates. We will use here two of them, one due to Chandrasekhar \cite{chandra:1986} and the other due to Kompaneets \cite{Kompaneets:1958} and Jordan, Ehlers and Kundt \cite{Jordan:1960}, passing from one to the other when convenient.

\subsection{Chandrasekhar's form of the most general cylindrically symmetric line element}

Following Ref. \cite{chandra:1986}, the most general cylindrically symmetric line element in cylindrical-like coordinates $x^\alpha=(t,\rho,\phi,z)$ reads 
\beq\fl\quad
\label{non_diag}
ds^2= e^{2\nu}(-dt^2+d\rho^2)+\rho \left[\chi d\phi^2 +\frac{1}{\chi}(dz-q_2 d\phi)^2\right]
\equiv ds^2_{(t,\rho)}+ds^2_{(\phi,z)}\,,
\eeq
where
\begin{eqnarray}
ds^2_{(t,\rho)}&=&g_{(t,\rho)}{}_{ab}dx^a dx^b
= e^{2\nu}(-dt^2+d\rho^2)\,,\nonumber\\
ds^2_{(\phi,z)}&=&g_{(\phi,z)}{}_{AB}dx^A dx^B
=\rho \left[\chi d\phi^2 +\frac{1}{\chi}(dz-q_2 d\phi)^2\right]\,,
\end{eqnarray}
with $x^a=(t,\rho)$, $x^A=(\phi,z)$, and all metric functions depend on $x^a$ only.
The determinant of the 2-metric  $g_{(\phi,z)}$ is
${\rm det}\left[g_{(\phi,z)}\right] = \rho^2$, 
and  gives an intrinsic, geometric definition to the coordinate $\rho$. Moreover
\beq
\label{area}
A=\sqrt{{\rm det}\left[g_{(\phi,z)}\right]} 
=\sqrt{|\partial_z|^2 |\partial_\phi|^2-(\partial_z\cdot \partial_\phi)^2}
= \rho \,,
\eeq
denotes the area (per unit proper length along the symmetry axis) spanned by the Killing vectors $\partial_\phi$ and $\partial_z$ associated with the cylindrical symmetry.\footnote{
This area element has the dimensions of a length and is often written as $2\pi \rho$. 
}

The vacuum Einstein's equations for the metric \eqref{non_diag} read
\begin{eqnarray}
\label{set1}
 \left( \rho\frac{\chi_t}{\chi}\right)_t- \left(\rho \frac{\chi_\rho}{\chi} \right)_\rho &=&-\frac{\rho}{\chi^2}[(q_{2,t})^2-(q_{2,\rho})^2]\,,\nonumber\\
 \left(\frac{\rho}{\chi^2}q_{2,t} \right)_t -\left(\frac{\rho}{\chi^2}q_{2,\rho} \right)_\rho  &=&0\,,
\end{eqnarray}
and
\begin{eqnarray}
\label{set2}
 \nu_t &=& \frac{\rho}{2\chi^2}(\chi_t \chi_\rho+q_{2,t}q_{2,\rho})\,,\nonumber\\
  \nu_{\rho} &=&-\frac{1}{4\rho}+\frac{\rho}{4\chi^2}[(\chi_t)^2+(\chi_\rho)^2+(q_{2,t})^2+(q_{2,\rho})^2]\,,
\end{eqnarray}
with $\chi_t=\partial_t \chi$, etc.
The first set of equations, Eq. \eqref{set1}, is the fundamental set, in the sense that once a solution of these equations is found, the last two equations for $\nu$, Eq. \eqref{set2}, can be integrated by quadratures.
The nice feature of Eqs. \eqref{set1} and \eqref{set2} is that the compatibility conditions for the second set $\nu_{,t\rho}=\nu_{,\rho t}$ is the first set, or in other words, the $1$-form field $(d\nu)^\flat$ is closed, the symbol $\flat$ denoting the fully covariant form of a tensor.

It is convenient to define the two (real) potentials, $\Psi$ and $\Phi$, such that
\beq
\Psi =\frac{\rho}{\chi}\,,\qquad
\frac{\rho}{\chi^2}q_{2,\rho}=\Phi_t\,,\qquad 
\frac{\rho}{\chi^2}q_{2,t}=\Phi_\rho\,,
\eeq
which enter the Ernst (complex) potential
\beq
Z=\Psi+i \Phi\,,
\eeq
satisfying the Ernst equation
\beq
{\rm Re}(Z)\left[ Z_{tt}-\frac{1}{\rho} (\rho Z_\rho)_\rho \right]=Z_t^2-Z_\rho^2\,.
\eeq
An equivalent form uses the fractional linear combination
\beq
E=\frac{Z-1}{Z+1}\,,
\eeq
fulfilling the companion equation
\beq
(1-|E|^2)\left[ E_{tt}-\frac{1}{\rho} (\rho E_\rho)_\rho \right]=-2E^* (E_t^2-E_\rho^2)\,.
\eeq

Finally, we notice an equivalent form of the metric largely used in the literature introduced in Refs. \cite{Kompaneets:1958,Jordan:1960} (we shortly recall this form of the metric with associated Einstein equations in Appendix A).
The metric functions $\nu$, $\chi$ and $q_2$ of Eq. \eqref{non_diag} are replaced by $\gamma$, $\psi$ and $\omega$, such that
\beq
\label{newfun}
\nu=\gamma -\psi\,,\qquad 
\frac{\rho}{\chi}=e^{2\psi}\,, 
\qquad q_2=-\omega\,.
\eeq

\subsection{Adapted frames}

We choose as fiducial observers those at rest with respect to the coordinates, i.e., with four-velocity $u=e_0=e^{-\nu}\partial_t$.
A natural orthonormal frame $\{e_\alpha\}$ adapted to them is built up with the triad
\beq
\label{frame}
e_1=e^{-\nu}\partial_\rho\,,\quad 
e_2=\frac{1}{\sqrt{\rho\chi}}(\partial_\phi+q_2\partial_z)\,,\quad 
e_3=\sqrt{\frac{\chi}{\rho}}\partial_z\,.
\eeq
This frame is also a spacetime (degenerate) Frenet-Serret frame along any integral curve of $e_0$ with 
\beq
\label{FS}
\kappa=e^{-\nu}\nu_\rho\,,\qquad 
\tau_1=0\,, \qquad
\tau_2=\frac{e^{-\nu}}{2\chi}q_{2,t}\,,
\eeq
with $\kappa$ and $\tau_2$  functions of of the proper time parameter along the world lines of the congruence $e_0$ so that 
\beq\fl\quad
\nabla_{u}e_0=\kappa e_1\,,\qquad
\nabla_{u}e_1=\kappa e_0\,,\qquad
\nabla_{u}e_2=\tau_2 e_3\,,\qquad
\nabla_{u}e_3=-\tau_2 e_2\,.
\eeq
The evolution of the spatial triad, projected orthogonally to $u=e_0$ so to coincide with the Fermi-Walker temporal derivative along $u$ can be summarized by introducing the Fermi-Walker angular velocity vector\footnote{
When dealing with a single world line adapted frame the angular velocity is usually termed $\omega_{\rm FS}$. The use of $\omega_{({\rm fw},u,e)}$ is preferable for a field of adapted frames along a congruence.
}
\beq
\omega_{({\rm fw},u,e)}=\tau_2 e_1\,,
\eeq
implying \cite{Jantzen:1992rg}
\beq
\label{omegafwdef}
P(u)\nabla_{u}e_a\equiv \nabla_{\rm (fw)}e_a=\omega_{({\rm fw},u,e)}\times e_a\,,
\eeq
where the spatial covariant derivative $\nabla(u)=P(u)\nabla$ is obtained by projecting $\nabla$ onto the local rest space of $u$ using the spatial projector $P(u)^\alpha_\beta=\delta^\alpha_\beta+u^\alpha u_\beta$.

In the case of diagonal metrics $(q_2=0)$ the spatial triad \eqref{frame} is already a Fermi-Walker transported triad, since $\tau_2=0$ from Eq. \eqref{FS}.
In the general non-diagonal case a Fermi-Walker transported triad is obtained simply by rotating the frame vectors $e_a$ orthogonally to $e_1$ (i.e., orthogonally to the direction of $\omega_{({\rm fw},u,e)}$),
\beq
\label{fwframe}
\tilde e_2 = \cos \beta e_2 +\sin \beta e_3\,,\qquad
\tilde e_3 = -\sin \beta e_2 +\cos \beta e_3\,.
\eeq
Choosing $\beta$ such that
\beq
\partial_t \beta =e^{\nu}\tau_2=\frac{q_{2,t}}{2\chi}\,,
\eeq
implies that the spatial frame $\{e_1,\tilde e_2,\tilde e_3\}$ is Fermi-Walker transported along $u$.

Finally, let us summarize the kinematical properties of the congruence associated with the static observers, whose unit timelike vector $u$ is orthogonal to the $t=$constant hypersurfaces $u^\flat = -e^{\nu} dt\equiv-N dt$.
The lapse factor is then $N=e^{\nu}$, whereas the shift vector is identically vanishing.
The congruence is accelerated, with acceleration vector
\beq
a(u)=e^{-\nu}\nu_\rho e_1\,,
\eeq
expanding, with nonvanishing frame components of the expansion tensor (symmetric part of the projected covariant derivative)
\beq\fl\quad
\theta(u)_{11}=e^{-\nu}\nu_t\,,\qquad
\theta(u)_{22}=e^{-\nu}\frac{\chi_t}{2\chi}=-\theta(u)_{33}\,,\qquad
\theta(u)_{23}=e^{-\nu}\frac{q_{2t}}{2\chi}\,,
\eeq
and irrotational ($\omega(u)=0$, antisymmetric part of the projected covariant derivative).

The spatial geometry of the $t=$constant hypersurfaces is associated with the metric
\beq
{}^{(3)}ds^2= e^{2\nu}d\rho^2+ \rho  \chi d\phi^2 +\frac{\rho }{\chi}(dz-q_2 d\phi)^2\,.
\eeq
The spatial geometry of the $t=$constant and $\rho=$constant hypersurfaces, i.e., of the wave fronts $\phi-z$, is identically flat since all metric quantities depend only on $t$ and $\rho$.

With the frame \eqref{frame} one can form then a Newman-Penrose complex null tetrad in a standard way, 
\beq
l=\frac{1}{\sqrt{2}}(e_0+e_1)\,,\quad n=\frac{1}{\sqrt{2}}(e_0-e_1)\,,\quad m=\frac{1}{\sqrt{2}}(e_2+i e_3)\,.
\eeq
Computing the Weyl scalars one can evaluate the Petrov spectral type of these spacetime, which turns out to be type I, in general \cite{Stachel:1966}.

\section{Timelike geodesics}

The timelike geodesics with unit tangent vector $U=\frac{dx^\alpha}{d\tau}\partial_\alpha$ ($U\cdot U=-1$) have the following simplified form 
\begin{eqnarray}
\label{geosgen}
\frac{dt}{d\tau}&=&\left[\left(\frac{d\rho}{d\tau}\right)^2+e^{-2\nu}\left(1+\frac{\chi K_2^2}{\rho}+\frac{X^2}{\rho\chi}\right)\right]^{1/2}
\,,\nonumber\\
\frac{d^2\rho}{d\tau^2}&=&-\frac{e^{-2\nu}}{2\rho\chi^2}\chi_\rho(-X^2+K_2^2\chi^2)\nonumber\\
&&
-\nu_\rho\left[2\left(\frac{d\rho}{d\tau}\right)^2+e^{-2\nu}\left(1+\frac{\chi K_2^2}{\rho}+\frac{X^2}{\rho\chi}\right)\right]\nonumber\\
&&
-2\nu_t\frac{dt}{d\tau}\frac{d\rho}{d\tau}-e^{-2\nu}\frac{K_2X}{\rho\chi}q_{2\rho}+\frac12e^{-2\nu}\frac{X^2+K_2^2\chi^2}{\rho^2\chi}
\,,\nonumber\\
\frac{d\phi}{d\tau}&=&\frac{K_1+q_2K_2}{\rho\chi}
\equiv\frac{X}{\rho\chi}
\,,\nonumber\\
\frac{dz}{d\tau}&=&\frac{K_2\chi^2+q_2X}{\rho\chi}
\,,
\end{eqnarray}
due to the Killing symmetries 
\beq
U_\phi=K_1\,,\qquad
U_z=K_2\,.
\eeq
In the case of nondiagonal metrics ($q_2\not =0$) the motion is in general never confined to a $z=$ constant hypersurface (including the symmetry plane $z=0$), since one should satisfy
\beq
\label{equat_mot}
K_2(\chi^2+q_2^2)+K_1q_2 =0\,,
\eeq
which represents a very strong restriction on $\chi$ and $q_2$ provided that either $K_1\not=0$ or $K_2\not=0$.  
Motion confined to any $z=$ constant hyperplane is always possible, instead, in the special case of diagonal metric ($q_2=0$), with $K_2=0$ and $X=K_1$ from Eq. \eqref{equat_mot}. In this case  the above equations simplify to
\begin{eqnarray}
\label{geosequat}
\frac{dt}{d\tau}&=&\left[\left(\frac{d\rho}{d\tau}\right)^2+e^{-2\nu}\left(1+\frac{K_1^2}{\rho\chi}\right)\right]^{1/2}
\,,\nonumber\\
\frac{d^2\rho}{d\tau^2}&=&\frac{e^{-2\nu}}{2\rho\chi^2}\chi_\rho K_1^2
-\nu_\rho\left[2\left(\frac{d\rho}{d\tau}\right)^2+e^{-2\nu}\left(1+\frac{K_1^2}{\rho\chi}\right)\right]\nonumber\\
&&
-2\nu_t\frac{dt}{d\tau}\frac{d\rho}{d\tau}+\frac12e^{-2\nu}\frac{K_1^2}{\rho^2\chi}
\,,\nonumber\\
\frac{d\phi}{d\tau}&=&\frac{K_1}{\rho\chi}
\,.
\end{eqnarray}

\subsection{Radial geodesics}

Setting $K_1=0=K_2$ (so that $X=0$ too) in Eq. \eqref{geosgen} gives the radial geodesics ($\phi=$ constant)
\begin{eqnarray}
\label{geosrad}
\frac{dt}{d\tau}&=&\left[\left(\frac{d\rho}{d\tau}\right)^2+e^{-2\nu}\right]^{1/2}
\,,\nonumber\\
\frac{d^2\rho}{d\tau^2}&=&
-\nu_\rho\left[2\left(\frac{d\rho}{d\tau}\right)^2+e^{-2\nu}\right]
-2\nu_t\frac{dt}{d\tau}\frac{d\rho}{d\tau}
\,.
\end{eqnarray}
Note that the last equation can be equivalently written in matrix form as
\beq
\frac{d^2\rho}{d\tau^2}=-\left(
\begin{array}{c}
\displaystyle \frac{dt}{d\tau} \,\,\displaystyle\frac{d\rho}{d\tau} 
\end{array}
\right)
{\mathcal A}
\left(
\begin{array}{c}
\displaystyle \frac{dt}{d\tau} \\
\displaystyle  \frac{d\rho}{d\tau} 
\end{array}
\right)\,,
\eeq
with ${\mathcal A}_{ab}=\Gamma^{\rho}{}_{ab}$, i.e.,
\beq\fl\quad
{\mathcal A}=\left(
\begin{array}{cc}
\nu_\rho & \nu_t \\
\nu_t & \nu_\rho 
\end{array}
\right)
=\nu_\rho {\rm Id}+\nu_t \sigma_1\,,\quad
{\rm Id}=\left(
\begin{array}{cc}
1 & 0 \\
0 & 1 
\end{array}
\right)\,,\qquad
\sigma_1=\left(
\begin{array}{cc}
0 & 1 \\
1 & 0 
\end{array}
\right)
\,,
\eeq
when expressed in terms of the Pauli matrices.

For later use it is convenient to cast Eqs. \eqref{geosrad} in an equivalent form using the frame representation of $U$
\beq
\label{Urad}
U=\Gamma(e_0+V e_1)\,,\qquad
\Gamma=(1-V^2)^{-1/2}\,,
\eeq
so that 
\beq
\label{georad1}
\frac{dV}{d\tau}=-\frac{e^{-\nu}}{\Gamma}(\nu_tV+\nu_\rho)\,, \qquad
\frac{dt}{d\tau}=\Gamma e^{-\nu}\,,\qquad
\frac{d\rho}{d\tau}=\Gamma V e^{-\nu}\,,
\eeq
where the linear velocity $V$ can be positive/negative corresponding to outward/inward radial motion.
Parametrizing the motion through the coordinate time $t$ instead of the proper time $\tau$ then yields
\beq
\label{georad2}
\frac{dV}{dt}=-(1-V^2)(\nu_tV+\nu_\rho)\,, \qquad
\frac{dt}{d\tau}=\Gamma e^{-\nu}\,,\qquad
\frac{d\rho}{dt}=V\,,
\eeq
which can be numerically integrated once a solution for the background gravitational field is specified.

\section{World line deviation and strains}

Consider a bunch of test particles at rest with respect to the coordinate system, i.e., a congruence ${\mathcal C}_u$ of timelike world lines, with unit tangent vector $u=e_0=e^{-\nu}\partial_t$.
The separation  between two neighboring particles, i.e., between a reference world line, say ${\mathcal C}_*$, and a generic curve of the congruence, is represented by a connecting vector $Y$, i.e., a vector undergoing Lie transport along $u$, $\nabla_uY=\nabla_Yu$.
Taking the covariant derivative along $u$ of both sides of the previous equation gives rise to the \lq\lq relative acceleration equation"
\beq
\label{eq:3bis}
\frac{D^2 Y^\alpha }{d \tau^2 }=-[E(u)^\alpha{}_\beta-
\nabla_\beta a(u)^\alpha]  Y^\beta\,,
\eeq
where $E(u)^\alpha{}_\gamma =R^\alpha{}_{\beta\gamma\delta}u^\beta u^\delta$ is the electric part of the Riemann tensor with respect to $u$ and represents the  \lq\lq tidal force" contribution to the relative deviation, whereas $\nabla_Y a(u)$ is the \lq\lq inertial"  contribution due to the observer's acceleration $a(u)=\nabla_u u$.
Deviation vector components may have strongly different behavior according to the preferred reference frame set up for their measurements.
The corresponding observer-dependent analysis can be found in Ref. \cite{Bini:2006vp,Bini:2007zzb}.

The relative deviation equation (\ref{eq:3bis}) with respect to this frame \eqref{frame} then becomes 
\beq
\label{eq:secder3bis}
\ddot Y^a +{\mathcal K}_{(u,e)}{}^a{}_b  Y^b=0\ ,  
\eeq
where ${\mathcal K}_{(u,e)}{}^a{}_b=[T_{({\rm fw},u,e)}-S(u)+E(u)]^a{}_b$ are the components of the \lq\lq deviation'' matrix ${\mathcal K}_{(u,e)}$, the overdot denoting proper time derivative. 
The strain tensor $S(u)$ is defined by
\beq
\label{eq:strainsdef}
S(u)_{ab}=\nabla(u)_b a(u)_a+a(u)_aa(u)_b\,,
\eeq
and depends only on the congruence $u$ and not on the transport properties of the chosen spatial triad $e_a$ along $u$, differently from the frame-dependent tensor $T$, which turns out to be given by
\begin{eqnarray}
\label{Tdef}
T_{({\rm fw},u,e)}{}^a{}_b&=&\delta^a_b \omega_{({\rm fw},u,e)}^2-\omega_{({\rm fw},u,e)}^a \omega_{({\rm fw},u,e)}{}_b -\epsilon^a{}_{bf}\dot \omega_{({\rm fw},u,e)}^f\nonumber \\
&& -2\epsilon^a{}_{fc}\omega_{({\rm fw},u,e)}^f K(u)^c{}_b\,. 
\end{eqnarray}
Here 
\beq
K(u)^\beta{}_\alpha=-P(u)^\beta{}_\nu P(u)_\alpha{}^\mu  \nabla_\mu u^\nu
\eeq
is the kinematical tensor, and $\omega_{({\rm fw},u,e)}$ represents the Fermi-Walker angular velocity \eqref{omegafwdef}, namely the angular velocity with which the spatial triad $e_a$ rotates with respect to a Fermi-Walker transported triad along $u$.

\section{C-energy and super-energy densities}

\subsection{C-energy density}

In 1965 K.S. Thorne \cite{thorne_65} introduced the concept of \lq\lq C-energy flux vector"  in spacetimes admitting a \lq\lq whole-cylinder symmetry,'' i.e., spacetimes which are invariant under rotation about the symmetry axis and translation along it, and reflection in any plane containing the symmetry axis or perpendicular to it ($\phi\to-\phi$, $z\to-z$). 
In this special case the most general metric is diagonal (with $q_2=0$ in Eq. \eqref{non_diag} or $\omega=0$ in Eq. \eqref{non_diag2}).

Thorne's definition can then be extended to the non-diagonal case in a straightforward manner \cite{chandra:1986}. 
The C-energy flux 4-vector $P_{\rm C}$ is defined as
\beq
\label{PCdef}
P_{\rm C}^\alpha = \eta^{\alpha\beta\gamma\delta} e_3{}_\gamma e_2{}_\delta \frac{{\sf E}_{,\beta}}{A}
=\eta^{\alpha\beta 32} \frac{{\sf E}_{,\beta}}{\rho}\,,
\eeq 
($\eta_{\delta\alpha\beta\gamma}=\sqrt{-g}\epsilon_{\delta\alpha\beta\gamma}$, $\epsilon_{0123}=1$) 
satisfying the conservation law $\nabla_\alpha P_{\rm C}^\alpha =0$, with $A=\rho$ from Eq. \eqref{area}.
Decomposing then $P_{\rm C}$ on the frame $\{e_{\alpha}\}$ gives
\beq
\label{PCu}
P_{\rm C}(u)=-\frac{{\sf E}_{,1}}{\rho}e_0+\frac{{\sf E}_{,0}}{\rho}e_1
\equiv-{\mathcal E}_{\rm C} e_0+ {\mathcal P}_{\rm C} e_1\,,
\eeq
where
\beq
{\mathcal E}_{\rm C}=P_{\rm C} \cdot e_0\,,\qquad 
{\mathcal P}_{\rm C}=P_{\rm C} \cdot e_1\,,
\eeq
so that
\begin{eqnarray}
\label{C_en_and_mom}
{\mathcal E}_{\rm C}&=&\frac{e^{-\nu}}{\rho}\partial_\rho {\sf E}
=\frac{e^{-(\gamma-\psi)}}{8}(|I|^2 +|O|^2 ) 
\,,\nonumber\\
{\mathcal P}_{\rm C}&=&\frac{e^{-\nu}}{\rho}\partial_t {\sf E}
=\frac{e^{-(\gamma-\psi)}}{8}(|I|^2-|O|^2) 
\,,
\end{eqnarray}
where the complex quantities $I$ and $O$ are related to the polarizations of the wave  \cite{Piran:1985dk,Halilsoy:1988vz} (see Eqs. \eqref{IOpmdef} and \eqref{IOcomplex} below).

The scalar function ${\sf E}$ is defined by  
\begin{eqnarray}
\label{Cendef}
{\sf E} &=&-\frac{1}{2}\ln \frac{g^{\alpha\beta}A _{,\alpha} A_{,\beta}}{ |\partial_z|^2 }
= \nu+\ln  \sqrt{\Psi}
=\gamma \,,
\end{eqnarray}
and is called \lq\lq C-energy,'' as it represents the total gravitational energy per unit length between the axis of symmetry and the cylindrical radius $\rho$ at time $t$.
It can be suggestively rewritten as
\beq
{\sf E}=\ln \sqrt{g_{zz}g_{\rho\rho}}\,,
\eeq
a simple expression involving the area element associated with the $t=$const., $\phi=$const. hypersurfaces.

Finally, let us note that Eqs. \eqref{C_en_and_mom} show that the 4-vector $P_{\rm C}$ is generally timelike (${\mathcal E}_{\rm C}\ge {\mathcal P}_{\rm C}$), namely the relative velocity of $P_{\rm C}$ with respect to $e_0$, $\nu(P_{\rm C},e_0)$ is given by
\beq
\label{nuCdef}
|\nu(P_{\rm C},e_0)|=\bigg|\frac{{\mathcal P}_{\rm C}}{{\mathcal E}_{\rm C}}\bigg|=\bigg|\frac{|I|^2-|O|^2}{|I|^2+|O|^2}\bigg|\le 1\,.
\eeq
It becomes a null vector when either $|I|=0$ or $|O|=0$. The case $|I|=|O|$ implies instead that $P_{\rm C}$ is aligned with $e_0$.

\subsection{Bel-Robinson super-energy density} 

The Bel and Bel-Robinson tensors are built with the Riemann tensor and the Weyl tensor, respectively, and are both divergence-free in vacuum spacetimes, where they coincide. 
Their mathematical properties are reviewed, e.g., in Refs. \cite{Bonilla:1997,Senovilla:1999xz}.
Since we are interested in vacuum solutions, we will consider the following gravitational (Bel-Robinson) super-energy tensor
\beq
\label{BelR}
 T^{\rm (g)}{}_{\alpha\beta} {}^{\gamma\delta}
 =
    C_{\alpha\rho\beta\sigma} C^{\gamma\rho\delta\sigma}
   + {}^* C_{\alpha\rho\beta\sigma} {}^* C^{\gamma\rho\delta\sigma} 
\ ,
\eeq
defined in terms of the Weyl tensor $C_{\alpha\beta\gamma\delta}$ and its dual. 

In close analogy with electromagnetism one can introduce the following super-energy density and super-momentum density (or super-Poynting, spatial) vector \cite{bel58,bel59,bel62,rob59} with respect to a given observer with 4-velocity $u$ (such that $u\cdot u=-1$, $u$ unitary, timelike and future-pointing, with $P(u)=g+u\otimes u$ the projection map  orthogonally to $u$)
\begin{eqnarray}
\label{Belsuperendef}
 \mathcal{E}^{\rm{(g)}}(u)
 &=&  T^{\rm (g)}_{\alpha\beta\gamma\delta} u^\alpha u^\beta u^\gamma u^\delta\ ,
\nonumber\\
 \mathcal{P}^{\rm{(g)}}(u)_\alpha
 &=& P(u)^\epsilon{}_\alpha T^{\rm (g)}_{\epsilon\beta\gamma\delta} u^\beta u^\gamma u^\delta
\,.
\end{eqnarray}
The corresponding $1+3$ splitted version in terms of the electric ($E(u)$) and magnetic ($H(u)$) parts of the Weyl tensor is 
\begin{eqnarray}
\label{BelRsuperendef}
 \mathcal{E}^{\rm{(g)}}(u)
 &=&  {\rm Tr}\, [ E(u)\cdot E(u) + H(u)\cdot H(u) ]\ ,\nonumber\\
 \mathcal{P}^{\rm{(g)}} (u)_\alpha
 &=&  2[ E(u)\times_u H(u)]_\alpha\,,
\end{eqnarray}
with $E(u)$ and $H(u)$ defined by 
 \beq
\label{EHweyl}
E(u)_{\alpha\beta}=C_{\alpha\mu\beta\nu}u^\mu u^\nu\,,\qquad 
H(u)_{\alpha\beta}={}^*C_{\alpha\mu\beta\nu}u^\mu u^\nu\,.
 \eeq
Furthermore, $[A\times_u B]_{\alpha} = \eta (u)_{\alpha\beta\gamma} A^{\beta} {}_{\delta}\, B^{\delta\gamma}$ defines a spatial cross product for two symmetric spatial tensor fields ($A, B$), with $\eta (u)_{\alpha\beta\gamma} 
= u^\delta \,\eta_{\delta\alpha\beta\gamma}$ the unit volume 3-form.
The nonvanishing frame components of the electric and magnetic parts of the Weyl tensor for cylindrically symmetric spacetimes are listed in Appendix A.

Finally, the following \lq\lq gravitational" 4-vector \cite{Maartens:1997fg,Senovilla:1999xz}
\beq
\label{Pgu}
{\sf P}^{\rm (g)}(u)=  -\mathcal{E}^{(\rm{g})}(u) u +\mathcal{P}^{(\rm{g})}(u)\equiv [T^{\rm (g)}{}_{\epsilon\beta\gamma\delta} u^\beta u^\gamma u^\delta]\,,
\eeq
can then be naturally defined.

\section{Accelerated motion}

Long ago Poynting \cite{Poy} and Robertson \cite{Rob} discussed the effect on the motion of small bodies orbiting a star induced by the radiation emitted by the star itself. From the mathematical point of view, this problem was modeled by assuming a Thomson-type interaction between massive test particles and a surrounding test radiation field superposed to the gravitational background of a compact object, leading to a non-geodesic motion driven by an external force.
Denoting by $T^{\mu\nu}$ the (conserved, $T^{\mu\nu}{}_{;\nu}=0$) energy-momentum tensor associated with the radiation field and by $U$ the 4-velocity of the particle, such a drag force was written in the form
\beq
\label{PRforce}
{\mathcal F}_{\rm PR}(U)^\alpha= -\sigma P(U)^{\alpha\mu}T_{\mu\nu}U^\nu\,,
\eeq
where $P(U)=g+U\otimes U$ denotes the projection operator orthogonally to $U$, and $\sigma$ is the (constant) effective interaction cross section.

Let us consider a test particle moving along the radial direction in the spacetime of a cylindrical gravitational wave, i.e., with 4-velocity, Eq. \eqref{Urad},
\beq
U=\Gamma(e_0+Ve_1)\,.
\eeq
At the zeroth order approximation, the orbit is a geodesic satisfying the equations \eqref{georad1}.
Going beyond the geodesic motion, but still neglecting backreaction effects on the background metric, one can consider a Poynting-Robertson-like interaction between the particle with mass $m$ and the the gravitational energy flux associated with the wave, represented by either the C-energy 4-momentum $P_{\rm C}(u)$ (Eq. \eqref{PCu}) or the super-energy 4-momentum ${\sf P}^{\rm(g)}(u)$ (Eq. \eqref{Pgu}), which are both conserved quantities ($P_{\rm C}^\alpha{}_{;\alpha} =0$ and ${\sf P}^{\rm(g)}{}^\alpha{}_{;\alpha} =0$).
The equations of motion can then be written as 
\beq
\label{eqmotoaccC}
ma(U)={\mathcal F}_C(U)
\equiv-\sigma P(U)P_C\,,
\eeq 
thinking of the C-energy or
\beq
\label{eqmotoaccg}
ma(U)={\mathcal F}^{\rm(g)}(U)
\equiv-\sigma P(U){\sf P}^{\rm(g)}\,,
\eeq 
thinking of the super-energy, where now $\sigma$ (having a different physical meaning and different dimensions with respect the one in Eq. \eqref{PRforce}) can be interpreted as a coupling constant representing the strength of the energy-momentum transfer from the wave to the particle.
For the present analysis, $\sigma$ is treated as a free parameter.
The particle's 4-acceleration is given by
\beq
a(U)=\Gamma\left[ e^{-\nu}(\nu_\rho+V\nu_t)+\Gamma\frac{dV}{d\tau}\right]E_1\,,
\eeq
with 
\beq
E_1=\Gamma(Ve_0+e_1)\,,  
\eeq 
the unit spatial vector orthogonal to $U$, whereas 
\beq
P(U)P=\Gamma(V{\mathcal E}+{\mathcal P})E_1\,,
\eeq
in both cases $P=P_C$ and $P={\sf P}^{\rm(g)}$.

The equations of motion \eqref{eqmotoaccC} and \eqref{eqmotoaccg} then imply the following evolution equation for the linear velocity
\beq
\Gamma\frac{dV}{d\tau}=-V\left(\tilde\sigma{\mathcal E}+e^{-\nu}\nu_t\right)
-\tilde\sigma{\mathcal P}-e^{-\nu}\nu_\rho\,,
\eeq
where $\tilde\sigma=\sigma/m$ and all quantities are evaluated along the orbit with parametric equations
\beq
\frac{dt}{d\tau}=e^{-\nu}\Gamma\,,\qquad
\frac{d\rho}{d\tau}=e^{-\nu}\Gamma V\,.
\eeq
Using the coordinate time parametrization, the previous equations become
\beq
\frac{dV}{dt}=-\frac1{\Gamma^2}(V\nu_t+\nu_\rho)-\tilde\sigma\frac{e^{\nu}}{\Gamma^2}\left(V{\mathcal E}+{\mathcal P}\right)\,,
\eeq
and 
\beq
\frac{dt}{d\tau}=e^{-\nu}\Gamma\,,\qquad
\frac{d\rho}{dt}=V\,,
\eeq
which reduce to the geodesic equations \eqref{georad2} for $\tilde\sigma=0$.

\section{Explicit examples}

We will consider below three explicit solutions belonging to the class of cylindrically symmetric gravitational wave spacetimes: Chandrasekhar's solution for cylindrical monochromatic waves with two polarizations \cite{chandra:1986}, Einstein-Rosen waves \cite{Einstein:1937qu}, and gravitational wave pulses (Weber-Wheeler-Bonnor solution \cite{Weber:1957oib,Bonnor:1957}), both with a single polarization.

\subsection{Chandrasekhar's solution for monochromatic waves}

The spacetime metric associated with Chandrasekhar's solution \cite{chandra:1986} is given by Eq. \eqref{non_diag} with functions
\begin{eqnarray}
\Psi&=&\frac{1-F^2}{1+F^2-2F\cos\bar t}
\,,\nonumber\\
\nu&=& \frac12\ln(1+F^2-2F\cos\bar t)+K
\,,\nonumber\\
q_2&=&-\frac{2\bar\rho F_{\bar\rho}}{1-F^2}\cos\bar t-4\int_0^{\bar\rho}\frac{xF(x)^2}{(1-F(x)^2)^2} dx\,,
\end{eqnarray}
where $F=F(\bar\rho)$ and $K=K(\bar\rho)$, with $\bar\rho=\sigma\rho$ and $\bar t=\sigma t$.
The function $F$ satisfies the equation
\beq
F_{\bar\rho\bar\rho}+\frac{F_{\bar\rho}}{\bar\rho}+F= -\frac{2F(F_{\bar\rho}^2+F^2)}{1-F^2}  \,,
\eeq
with initial conditions $F=F_0$ ($0<F_0<1$) and $F_{\bar\rho}=0$ for $\bar\rho=0$, whereas $K$ follows by quadrature
\beq
K=K_0+\int_0^{\bar\rho}\left[\frac{x(F^2+F_x^2)}{(1-F^2)^2}+\frac{FF_x^2}{1-F^2}\right]dx\,.
\eeq
Approaching the axis ($\bar\rho\to0$) the solutions for $F$ and $K$ behave as
\begin{eqnarray}
F&=&F_0-\frac{F_0(1+F_0^2)}{4(1-F_0^2)}\bar\rho^2
+\frac{F_0(3F_0^2+1)(1+F_0^2)}{64(1-F_0^2)^2}\bar\rho^4
+O(\bar\rho^6)\,,\nonumber\\
K&=&K_0+\frac{F_0^2}{4(1-F_0^2)}\bar\rho^2
-\frac{F_0^2(1+F_0^2)}{64(1-F_0^2)^2}\bar\rho^4
+O(\bar\rho^6)\,.
\end{eqnarray}
In the limit $F_0\to0$ the above relations imply the following link with the Bessel functions
\begin{eqnarray}\fl
F&=&F_0\left(1-\frac14\bar\rho^2+\frac1{64}\bar\rho^4-\frac1{2304}\bar\rho^6+O(\bar\rho^8,F_0^2)\right)
=F_0J_0(\bar\rho)+O(F_0^2)\,,\nonumber\\
\fl
K&=&K_0+F_0^2\left(
\frac14\bar\rho^2-\frac1{64}\bar\rho^4+\frac1{1152}\bar\rho^6+O(\bar\rho^8,F_0^2)
\right)\nonumber\\
\fl
&=&K_0+F_0^2\left[\frac12\left(J_0(\bar\rho)^2-1\right)
-\bar\rho J_0(\bar\rho)J_1(\bar\rho)
+\bar\rho^2(J_0(\bar\rho)^2+J_1(\bar\rho)^2)
\right]
+O(F_0^4)\,,
\end{eqnarray}
where we have used the notation $J_0(\bar\rho)={\rm BesselJ}(0,\bar\rho)$ and $J_1(\bar\rho)={\rm BesselJ}(1,\bar\rho)$ for the Bessel functions of the first kind with index 0 and 1, respectively.
The first term in the $F_0$-expansion of $F(\bar\rho)$ has been found by Chandrasekhar \cite{chandra:1986}. 
Going to the next order, we find
\beq
F(\bar\rho)=F_0 \left[  J_0(\bar\rho) + F_0^2 F_2(\bar\rho)
\right]+O(F_0^5)\,,
\eeq
with
\begin{eqnarray}
F_2(\bar\rho) &=&\pi [J_0(\bar\rho) I_1(\bar\rho) +Y_0(\bar\rho) I_2(\bar\rho)] \nonumber\\
&=&-\frac12 \bar\rho^2+\frac{3}{32}\bar\rho^4-\frac{5}{576}\bar\rho^6+O(\bar\rho^{8})\,,
\end{eqnarray}
where $Y_0(\bar\rho)={\rm BesselY}(0,\bar\rho)$ is the Bessel function of the second kind with index $0$, and
\begin{eqnarray}
I_1(\bar\rho) &=&\int_0^{\bar\rho} x  Y_0(x)J_0(x)[J_0^2(x)+J_1^2(x)] dx\,, \nonumber\\
I_2(\bar\rho) &=&  \int_0^{\bar\rho} x  J_0^2(x)[J_0^2(x)+J_1^2(x)] dx\,.
\end{eqnarray}
The asymptotic behavior for $F$ at spatial infinity $\bar\rho\to\infty$ is instead given by \cite{chandra:1986} 
\begin{eqnarray}
F&=& \frac1{4\bar\rho^{1/2}}\cos\left(\bar\rho+\frac1{16}\ln\bar\rho+b\right)+O(\bar\rho^{-3/2})
\,,
\end{eqnarray}
with $b$ a constant, whereas $K$ grows linearly with $\bar\rho$ ($K\sim\bar\rho/16$).
Following Ref. \cite{chandra:1986} this is conveniently shown by expressing $F=\tanh(f/4)$, with $f$ satisfying the equation
\beq
\label{eqf}
f_{\bar\rho\bar\rho}+\frac{1}{\bar\rho}f_{\bar\rho}+\sinh f=0\,,
\eeq
which can be solved perturbatively.\footnote{
It is interesting to note that Eq. \eqref{eqf} can be obtained as an Euler-Lagrange equation associated with the Lagrangian density ${\mathcal L}(f_{\bar\rho},f;\bar\rho)=\bar\rho(f_{\bar\rho}^2-2\cosh f)$, namely $\frac{d}{d\bar\rho}\frac{\partial{\mathcal L}}{\partial f_{\bar\rho}}-\frac{\partial{\mathcal L}}{\partial f}=0$. 
}

The C-energy density \eqref{C_en_and_mom} turns out to be
\beq\fl\quad
{\mathcal E}_{\rm C}=\frac{\mathcal W}{\left[1-\frac{2F}{(1+F)^2}(1-\cos\bar t)\right]^{1/2}}\,,\qquad
{\mathcal W}=\frac{e^{-K}\bar\rho(F^2+F_{\bar\rho}^2)}{(1-F^2)^2(1+F)}\,,
\eeq
and admits the following representation
\beq\fl\quad
{\mathcal E}_{\rm C}=\frac{{\mathcal W}(1+F)}{\left[ 1+F^2+ 2F \cos\bar t\right]^{1/2}}
={\mathcal W}(1+F) \sum_{n=0}^\infty P_n(-\cos\bar t)F^n\,,
\eeq
where $P_n(x)$ are the Legendre polynomials of order $n$.
The spatial C-momentum density ${\mathcal P}_{\rm C}$, instead, is identically vanishing, implying that $|I|=|O|$ (see Eq. \eqref{nuCdef}).
The C-energy scalar ${\sf E}$ is independent of time, as from Eq. \eqref{Cendef}. It reads
\beq
{\sf E}=K+\frac12\ln(1-F^2)\,.
\eeq
Furthermore, averaging ${\mathcal E}_{\rm C}$ over a period gives
\beq
\label{ell_fun}
\langle {\mathcal E}_{\rm C}\rangle\equiv\frac1{2\pi}\int_0^{2\pi}{\mathcal E}_{\rm C} d\bar t
={\mathcal W}\,\frac{2}{\pi}\,{\rm EllipticK}\left(\frac{2\sqrt{F}}{1+F}\right)\,,
\eeq
where 
\beq
{\rm EllipticK}(k)=\int_0^1\frac{dx}{\sqrt{(1-x^2)(1-k^2x^2)}}
\eeq
denotes the complete Elliptic integral of the first kind. In fact, defining
\beq
c_n = \frac{1}{2\pi}\int_0^{2\pi} P_n(-\cos\bar t) d\bar t
\eeq
we find $c_{2n+1}=0$, $n=0,1,\ldots$, and
\beq
c_0=1 \,,\quad c_2=\frac14 \,,\quad c_4=\frac{9}{64}\,,\quad c_6=\frac{25}{256}\,,
\ldots  
\eeq
Inserting these values into the definition \eqref{ell_fun} of $\langle {\mathcal E}_{\rm C}\rangle$ gives
\beq
\langle {\mathcal E}_{\rm C}\rangle = {\mathcal W}(1+F) \sum_{n=0}^\infty c_n F^n\,,
\eeq
with
\beq\fl\quad
\sum_{n=0}^\infty c_n F^n =1+\frac14 F^2+\frac{9}{64}F^4+O(F^6)
=\frac{2}{\pi (1+F)}\,{\rm EllipticK}\left(\frac{2\sqrt{F}}{1+F}\right)\,.
\eeq

\begin{figure}
\[
\begin{array}{cc}
\includegraphics[scale=0.25]{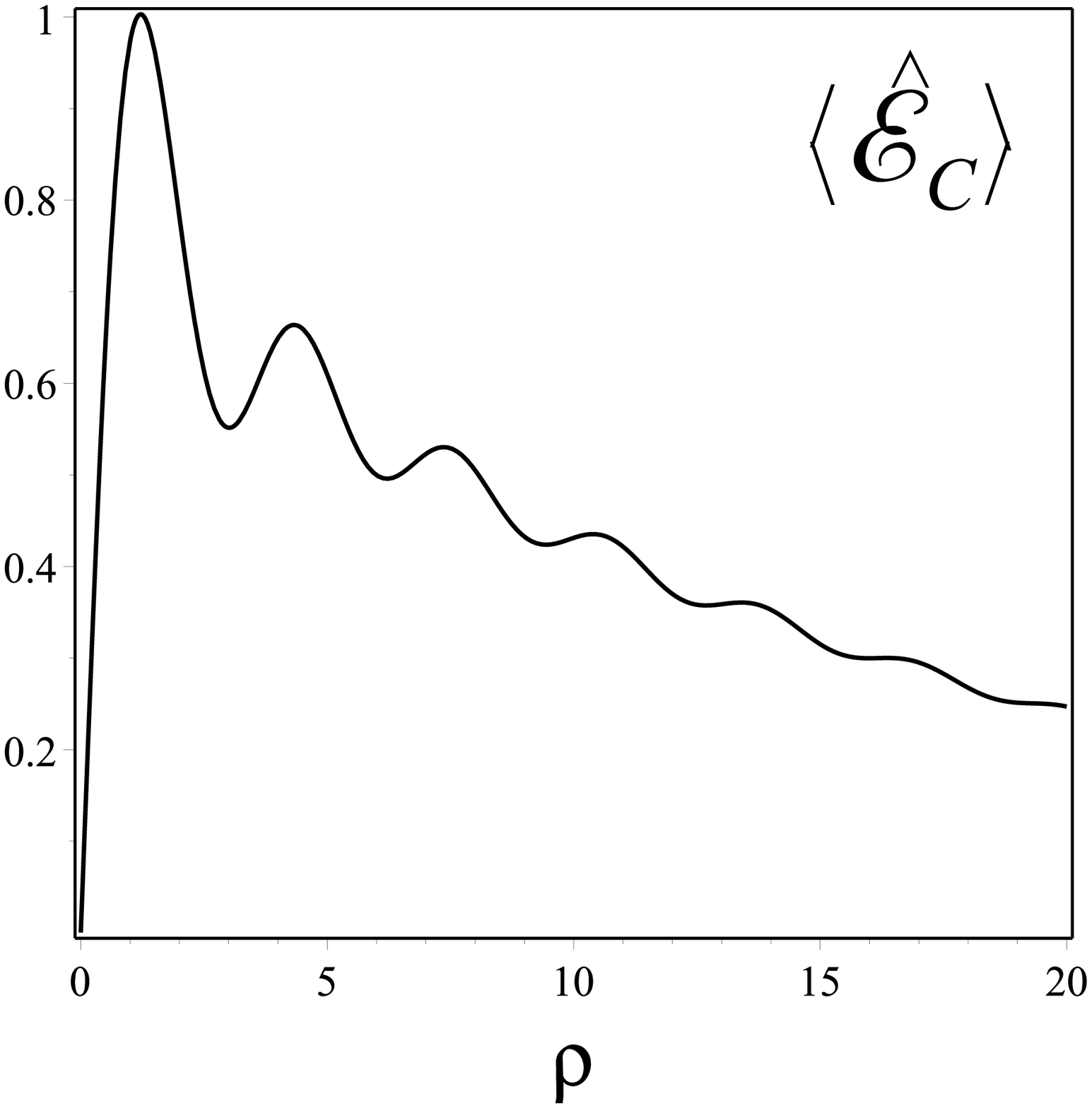}\qquad &  \includegraphics[scale=0.25]{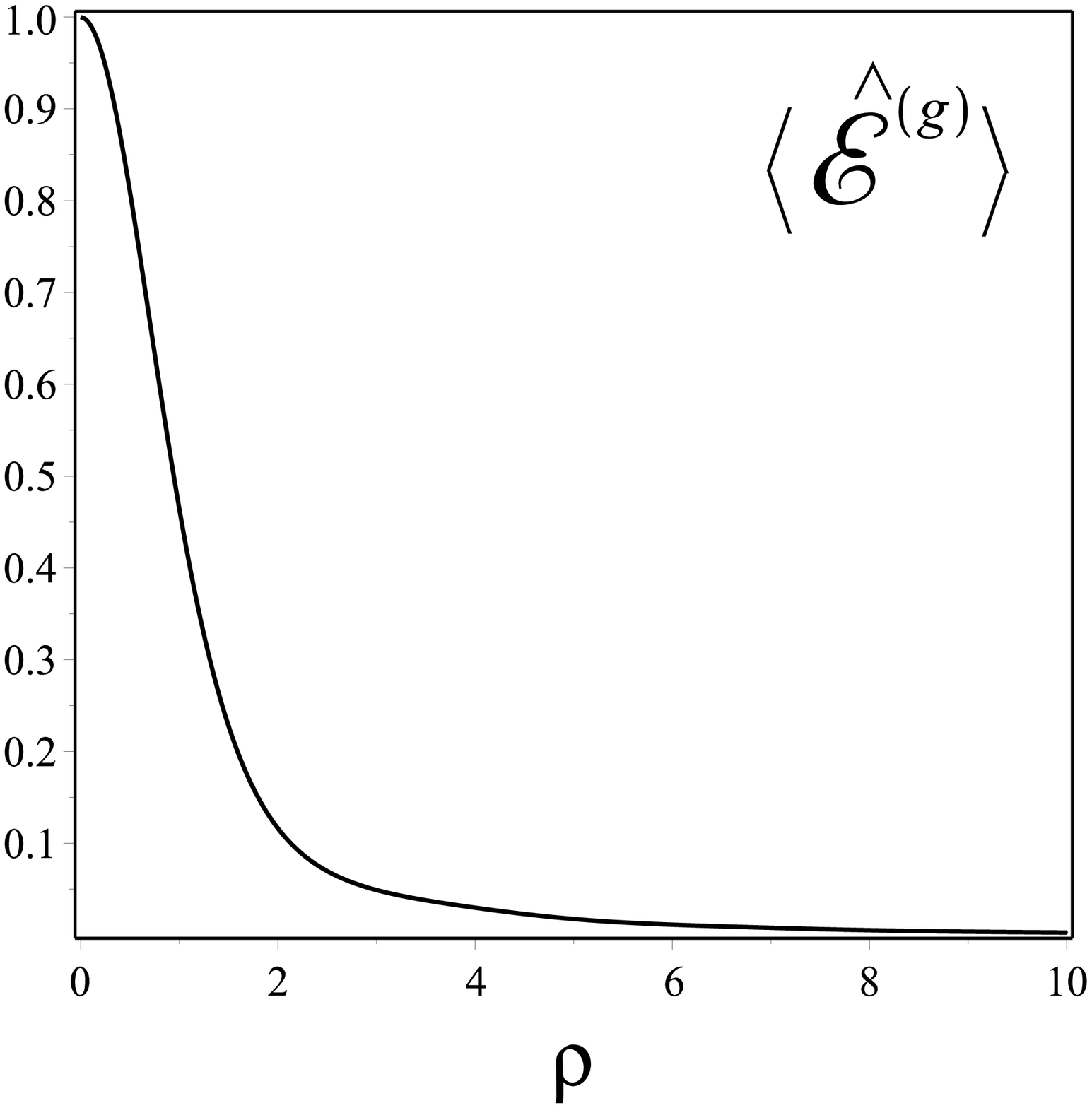}\cr
(a) & (b) 
\end{array}
\]
\caption{\label{fig:1} Chandrasekhar's solution. The behavior of the time-averaged (normalized, in order to highlight the comparison) C-energy $\langle \hat {\mathcal E}_{\rm C}\rangle=\langle {\mathcal E}_{\rm C}\rangle/\langle {\mathcal E}_{\rm C}\rangle_{\rm max}$ and super-energy $\langle \hat {\mathcal E}^{\rm(g)}\rangle=\langle {\mathcal E}^{\rm(g)}\rangle/\langle {\mathcal E}^{\rm(g)}\rangle_{\rm max}$ densities as functions of $\rho$ is shown in panels (a) and (b), respectively.
The parameters entering the solution are chosen as $\sigma=1$, $F_0=0.3$, $K_0=0$.
The corresponding maximum values are $\langle {\mathcal E}_{\rm C}\rangle_{\rm max}\approx 0.079$ and $\langle {\mathcal E}^{\rm(g)}\rangle_{\rm max}\approx 0.544$.}
\end{figure}

The expressions for the super-energy and super-momentum densities are straightforward to derive but quite involved, so we avoid showing them explicitly.
These quantities are in general both nonvanishing.
However, averaging over a period gives a vanishing super-momentum density.
The behavior of both time-averaged C-energy and super-energy densities as functions of $\rho$ is shown in Fig. \ref{fig:1}.
In order to highlight the comparison, the latter have been normalized by their maximum values, i.e., we have defined
$\langle \hat {\mathcal E}_{\rm C}\rangle=\langle {\mathcal E}_{\rm C}\rangle/\langle {\mathcal E}_{\rm C}\rangle_{\rm max}$ and $\langle \hat {\mathcal E}^{\rm(g)}\rangle=\langle {\mathcal E}^{\rm(g)}\rangle/\langle {\mathcal E}^{\rm(g)}\rangle_{\rm max}$.
We see that $\langle \hat {\mathcal E}_{\rm C}\rangle$ vanishes at the axis $\rho=0$, whereas $\langle \hat {\mathcal E}^{\rm(g)}\rangle$ is maximum there. 

\begin{figure}
\centering
\includegraphics[scale=0.27]{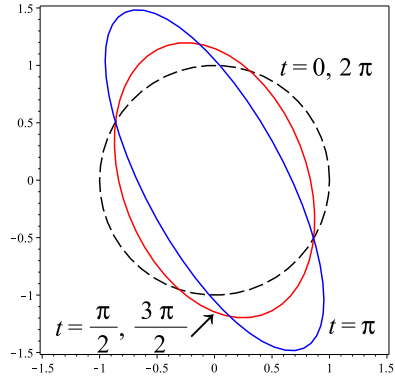}
\caption{\label{fig:2} Chandrasekhar's solution. 
Deformation of a ring of particles initially at rest at $\rho=4$ induced by the travelling wave at different coordinate times $t=[0,\frac{\pi}2,\pi,\frac{3\pi}2,2\pi]$, with the same parameter choice as in Fig. \ref{fig:1}.
The parameters entering the solution are chosen as $\sigma=1$, $F_0=0.3$, $K_0=0$.
}
\end{figure}

Fig. \ref{fig:2} shows the deformation of a ring of particles (initially at rest at a given radius in a plane orthogonal to the direction of propagation of the wave) induced by the wave at different coordinate times. 
The components of the deviation vector have been computed with respect to the Fermi-Walker spatial frame $\{e_1,\tilde e_2,\tilde e_3\}$ (see Eq. \eqref{fwframe}), so that the tensor $T_{({\rm fw},u,e)}$ vanishes identically in this case.
The deviation equation \eqref{eq:secder3bis} has been then integrated numerically to give the evolution of the shape in the plane $Y^2$-$Y^3$. One easily recognizes the effect of the polarizations in squeezing and rotating the shape of the bunch, as it happens in the well known weak field regime.
However, the present strong field situation implies a different (evolving) behavior, as expected. 

\begin{figure}
\[
\begin{array}{cc}
\includegraphics[scale=0.25]{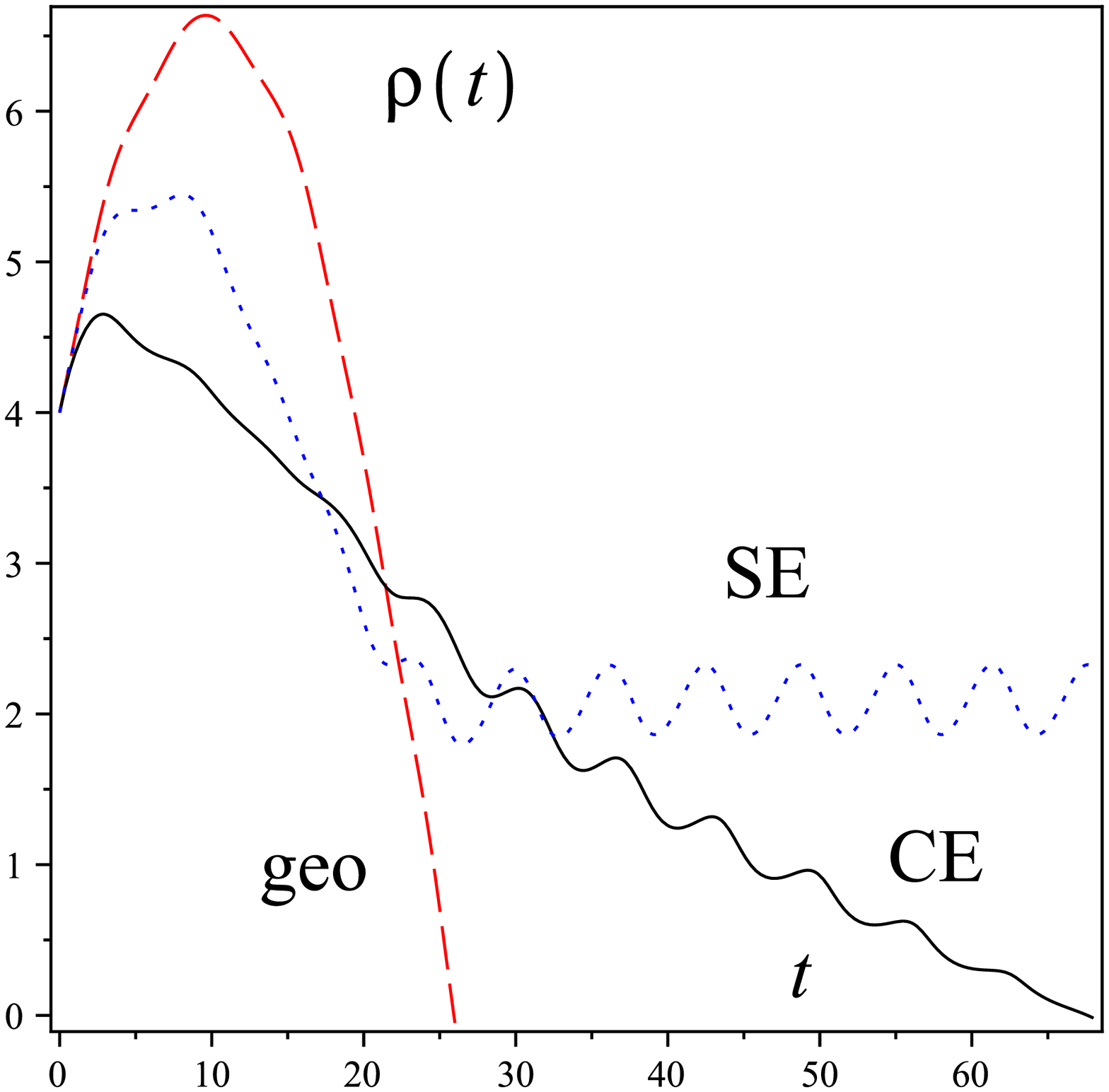}\qquad &  \includegraphics[scale=0.25]{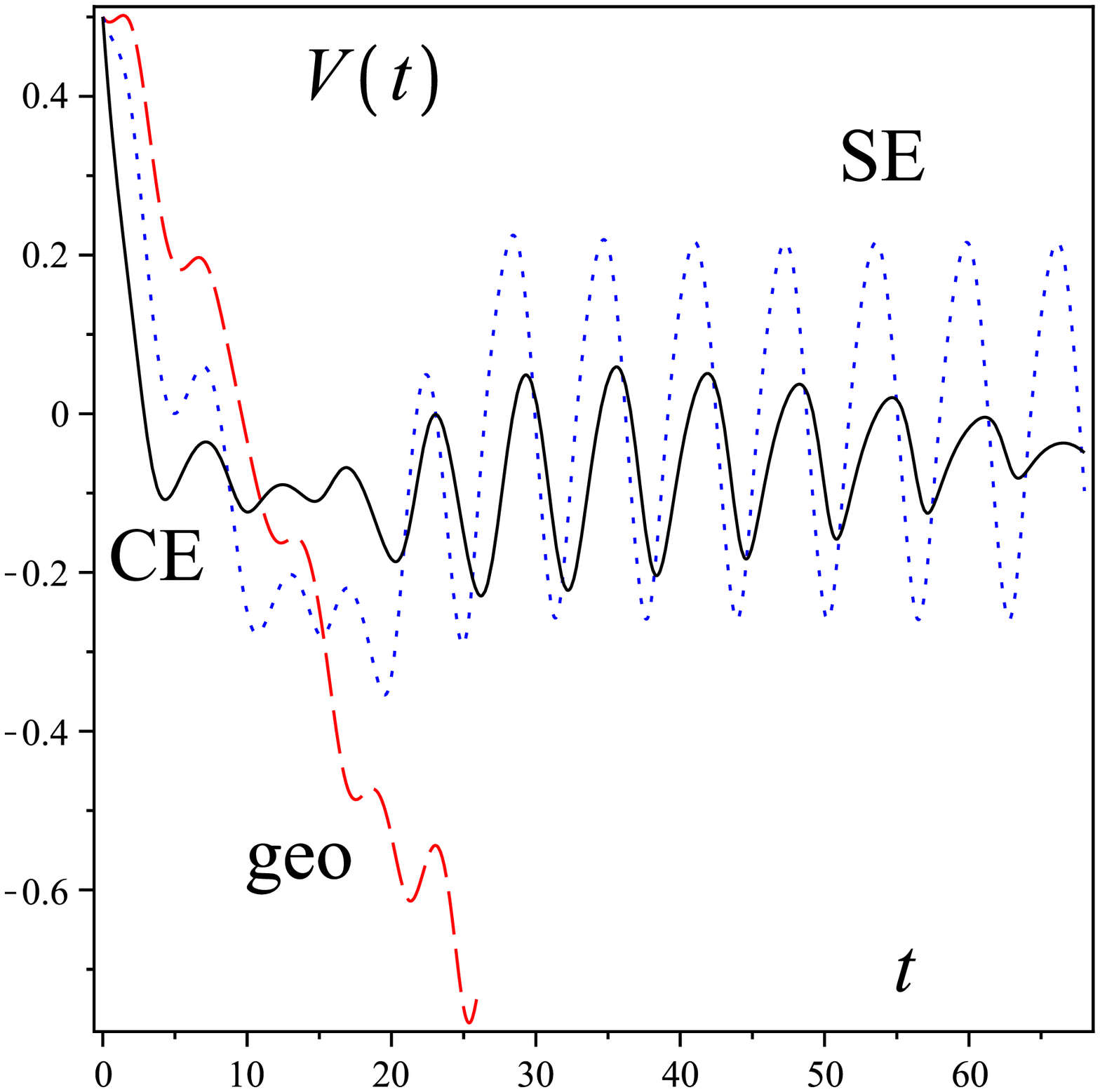}\cr
(a) & (b) \cr
\end{array}
\]
\caption{\label{fig:3} Chandrasekhar's solution. The equations for both geodesic (geo) and accelerated radial orbits due to both C-energy (CE) and super-energy (SE) energy-momentum transfer are numerically integrated for the same parameters as in Fig. \ref{fig:1} and initial conditions $\rho(0)=4$ and $V(0)=0.5$. The additional parameter $\tilde \sigma$ is chosen as $\tilde\sigma=10$ to enhance the  external force contribution.
The radial position $\rho(t)$ (panel a) and radial velocity $V(t)$ (panel b) are plotted as functions of $t$. 
}
\end{figure}

The numerical integration of the radial equations of motion, either geodesic or accelerated by a Poynting-Robertson-like interaction force, is shown in Fig. \ref{fig:3}.
The initial conditions have been chosen so that the particle starts moving outward from a given radius with a positive value of the linear velocity.
After an initial push forward, the wavy character of the solution forces it back towards the axis, which is reached after a few cycles in the geodesic case.
The effect of the interaction force (enhanced with the choice of a large value of the coupling parameter $\tilde\sigma$ for a better visualization in the plots) is to delay the turning back towards the axis, with a value of the linear velocity oscillating around a constant value close to zero.
This seems to be a general behavior for both C-energy and super-energy driven accelerations.
In the latter case an equilibrium solution $\rho=$constant seems to form due to the balance between purely gravitational (geodesic) effects due to the background and Poynting-Robertson-like (non-geodesic) effects.
This feature is reminiscent of the existence of equilibrium solutions (suspended orbits) occurring in the case of standard Poynting-Robertson effect in black hole spacetimes \cite{abram,ML,Bini:2008vk,Oh:2010qn,Bini:2011zza}.

% figure 1

% figure 2

% figure 3

\subsection{Einstein-Rosen waves}

\begin{figure}
\[
\begin{array}{cc}
\includegraphics[scale=0.25]{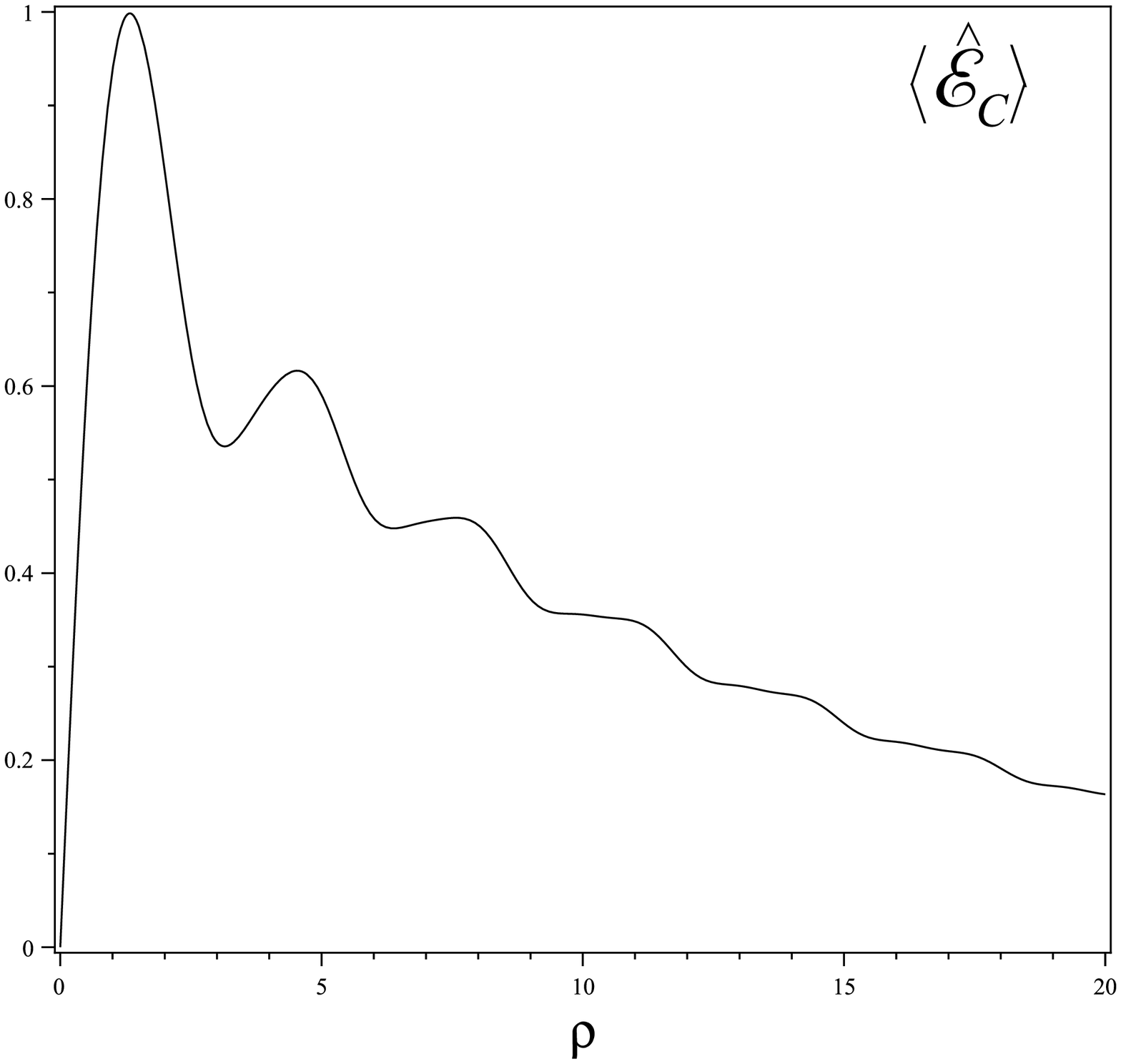}\qquad &  \includegraphics[scale=0.25]{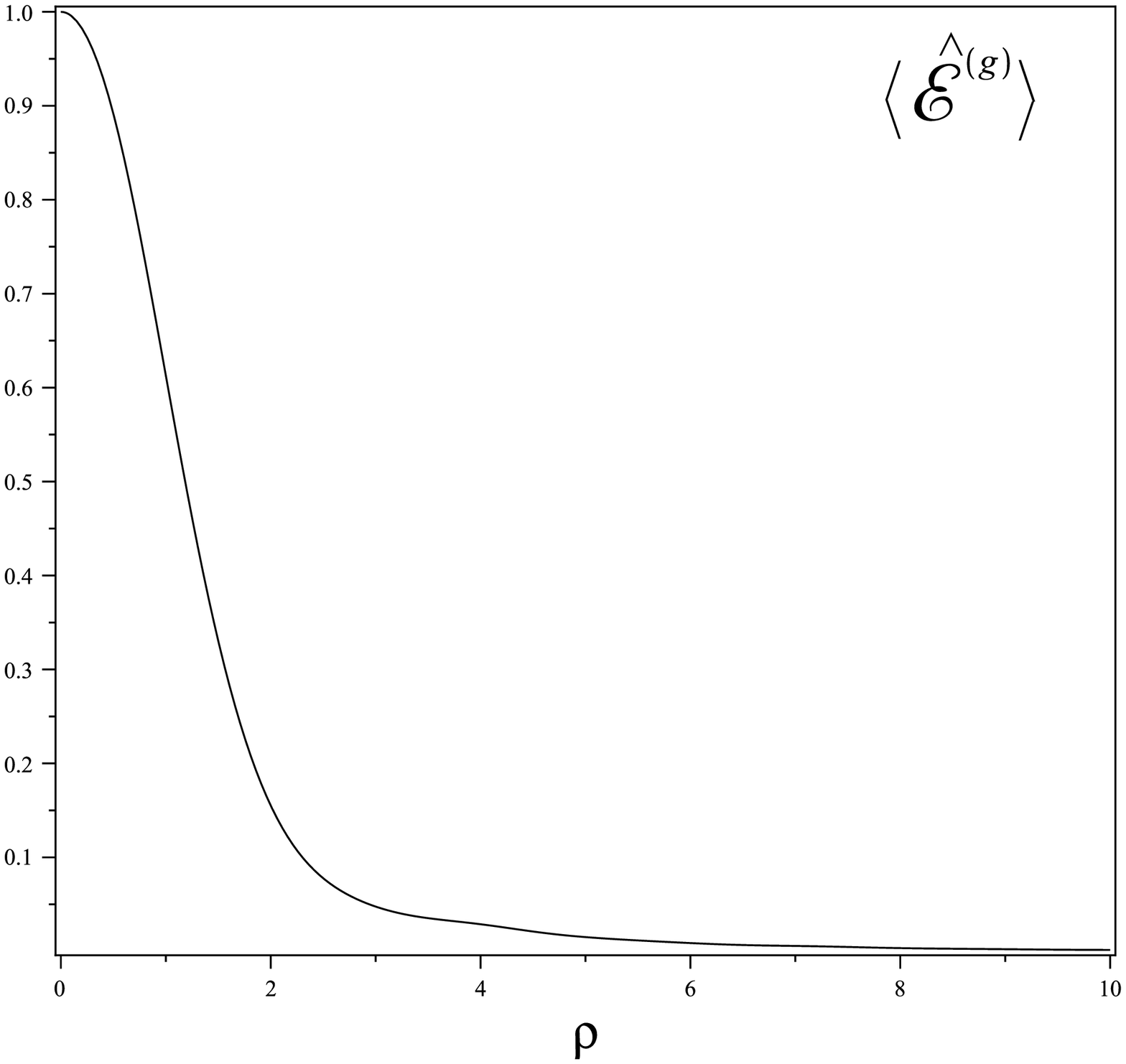}\cr
(a) & (b) 
\end{array}
\]
\caption{\label{fig:4} Einstein-Rosen waves. The behavior of the time-averaged (normalized) C-energy $\langle \hat {\mathcal E}_{\rm C}\rangle$ and super-energy $\langle \hat {\mathcal E}^{\rm(g)}\rangle$ densities as functions of $\rho$ is shown in panels (a) and (b), respectively.
The parameters entering the solution are chosen as $\sigma=1$, $A=1$.
The corresponding maximum values are $\langle {\mathcal E}_{\rm C}\rangle_{\rm max}\approx 0.998$ and $\langle {\mathcal E}^{\rm(g)}\rangle_{\rm max}\approx 0.492$.}
\end{figure}

\begin{figure}
\centering
\includegraphics[scale=0.27]{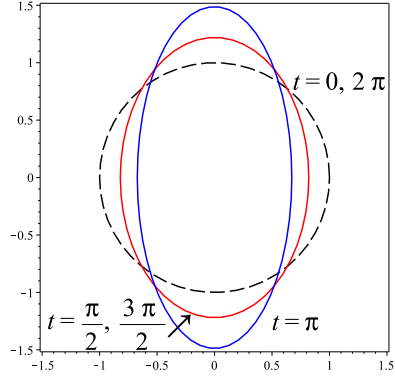}
\caption{\label{fig:5} Einstein-Rosen waves. 
Deformation of a ring of particles initially at $\rho=4$ induced by the travelling wave at different coordinate times $t=[0,\frac{\pi}2,\pi,\frac{3\pi}2,2\pi]$ and the same parameter choice as in Fig. \ref{fig:4}.
}
\end{figure}

\begin{figure}
\[
\begin{array}{cc}
\includegraphics[scale=0.25]{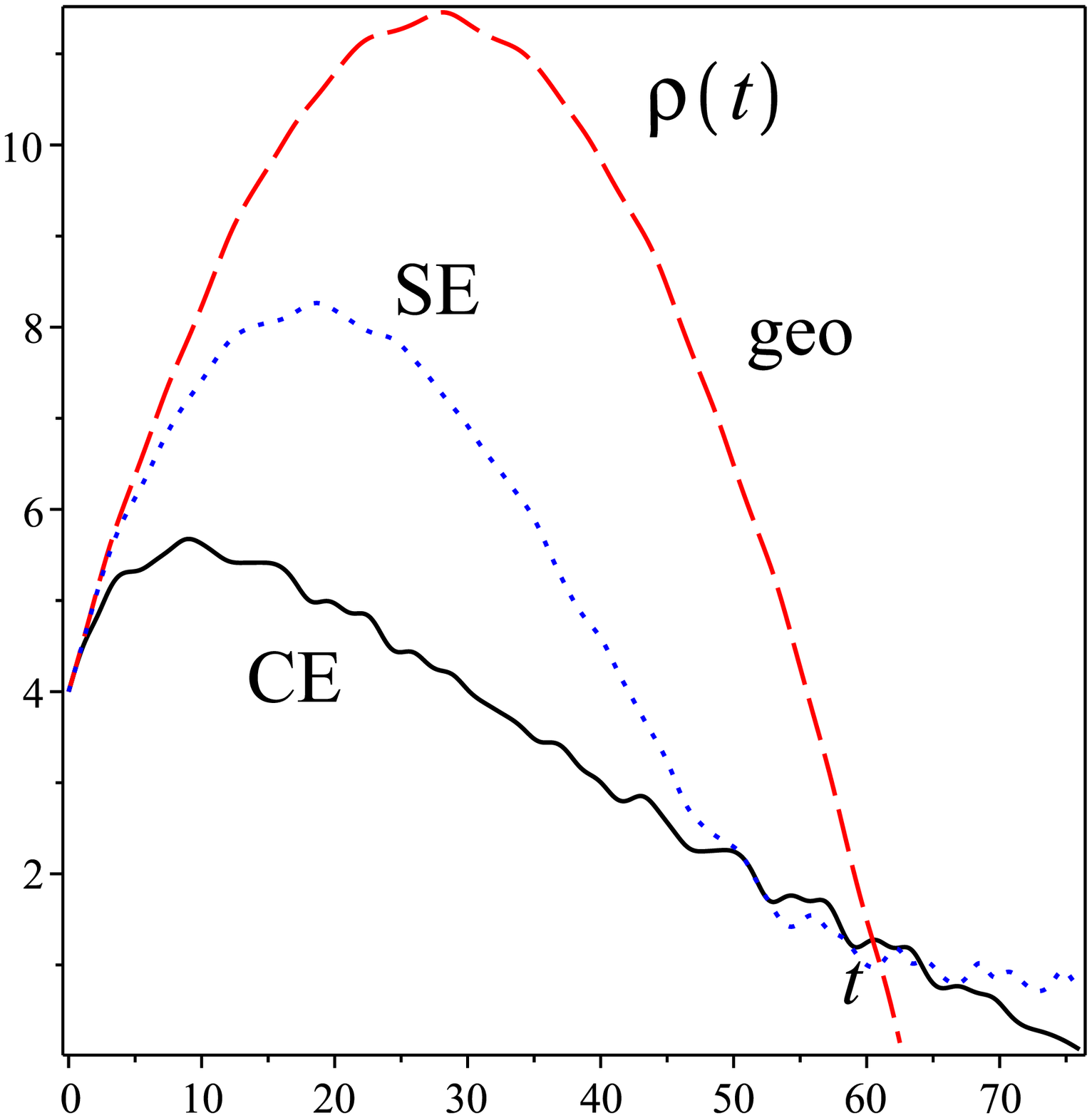}\qquad &  \includegraphics[scale=0.25]{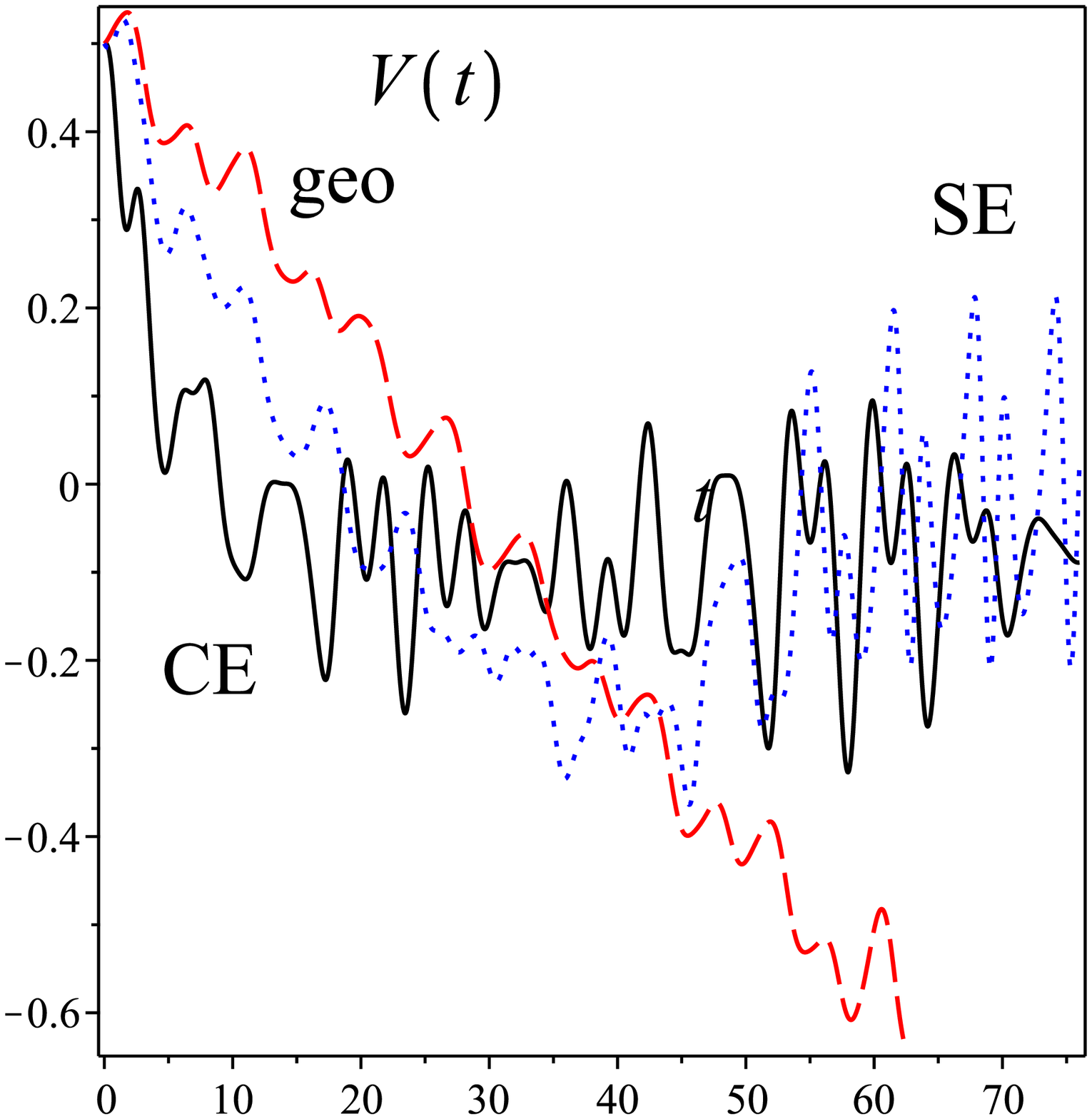}\cr
(a) & (b) 
\end{array}
\]
\caption{\label{fig:6} Einstein-Rosen waves. The equations for both geodesic and accelerated radial orbits due to both C-energy and super-energy energy-momentum transfer are numerically integrated for the same parameters as in Fig. \ref{fig:4} and initial conditions $\rho(0)=4$ and $V(0)=0.5$. The additional parameter $\tilde \sigma$ is chosen as $\tilde\sigma=10$ to enhance the  external force contribution.
The radial position $\rho(t)$ (panel a) and radial velocity $V(t)$ (panel b) are plotted as functions of $t$.
}
\end{figure}

The Einstein-Rosen spacetime \cite{Einstein:1937qu} has $q_2(t,\rho)=0$ and it is convenient to use $\Psi$ in place of $\chi$ in Eq. \eqref{non_diag} above.
The solution is given by
\begin{eqnarray}\fl\quad
\Psi&=& e^{A J_0(\bar\rho)\cos\bar t}\,,\nonumber\\
\fl\quad
\nu&=& -\frac12  AJ_0(\bar\rho)\cos\bar t
+\frac18 A^2 [\bar\rho^2 (J_0(\bar\rho)^2+J_1(\bar\rho)^2)
-2\bar\rho J_0(\bar\rho) J_1(\bar\rho)\cos^2\bar t]\,,
\end{eqnarray}
with $A$ dimensionless.
It exhibits similar features to those described above for Chandrasekhar's solution.
Fig. \ref{fig:4} shows the behavior of the time-averaged (normalized) C-energy and super-energy densities as functions of $\rho$.
The deformation of a ring of particles initially at rest is shown in Fig. \ref{fig:5}, the components of the deviation vector being computed with respect to the spatial frame \eqref{frame}, which is already a Fermi-Walker frame.
The absence of the cross polarization in this case explains the non-rotated shape.
Finally, the numerical integration of radial equations of motion is shown in Fig. \ref{fig:6}.

% figure 4

% figure 5

% figure 6

\subsection{Weber-Wheeler-Bonnor pulse}

The Weber-Wheeler-Bonnor solution \cite{Weber:1957oib,Bonnor:1957} describes a cylindrical pulse of gravitational radiation which is initially incoming and then reflects symmetrically off the axis.
The pulse is obtained as a linear superposition of Einstein-Rosen waves with a cut off in the frequency domain, i.e., 
\beq
\label{psi_def_WWB}
\psi=2C\int_0^\infty e^{-a\sigma}J_0(\sigma\rho)\cos\sigma t\, d\sigma\,,
\eeq
where $a$ and $C$ are constants, with $a$ determining the width of the pulse.\footnote{
The integral defining $\psi$ can be computed starting from
$$
{\mathcal J}=\int_0^\infty e^{i \sigma (t+ia)}J_0(\sigma\rho)\,  d\sigma\,.
$$ 
Combining the result for ${\mathcal J}$ and its complex conjugate one easily checks Eq. \eqref{psi_def_WWB}, also clarifying that 
in the final form of $\psi$ only enter the complex \lq\lq times" $t\pm ia$.
}
The metric is given by Eq. \eqref{non_diag2} with $\omega=0$ and 
\begin{eqnarray}
\psi &=& \sqrt{2}C\left[\frac{D_++a^2+\rho^2-t^2}{D_+^2}\right]^{1/2}
\,,\nonumber\\
\gamma &=&\frac{C^2}{2a^2}\left[1-\frac{2a^2\rho^2D_-^2}{D_+^4}+\frac{\rho^2-a^2-t^2}{D_+}\right]\,,
\end{eqnarray}
where 
\beq
D_\pm=\sqrt{(a^2+\rho^2-t^2)^2\pm4a^2t^2}\,.
\eeq

\begin{figure}
\[
\begin{array}{cc}
\includegraphics[scale=0.25]{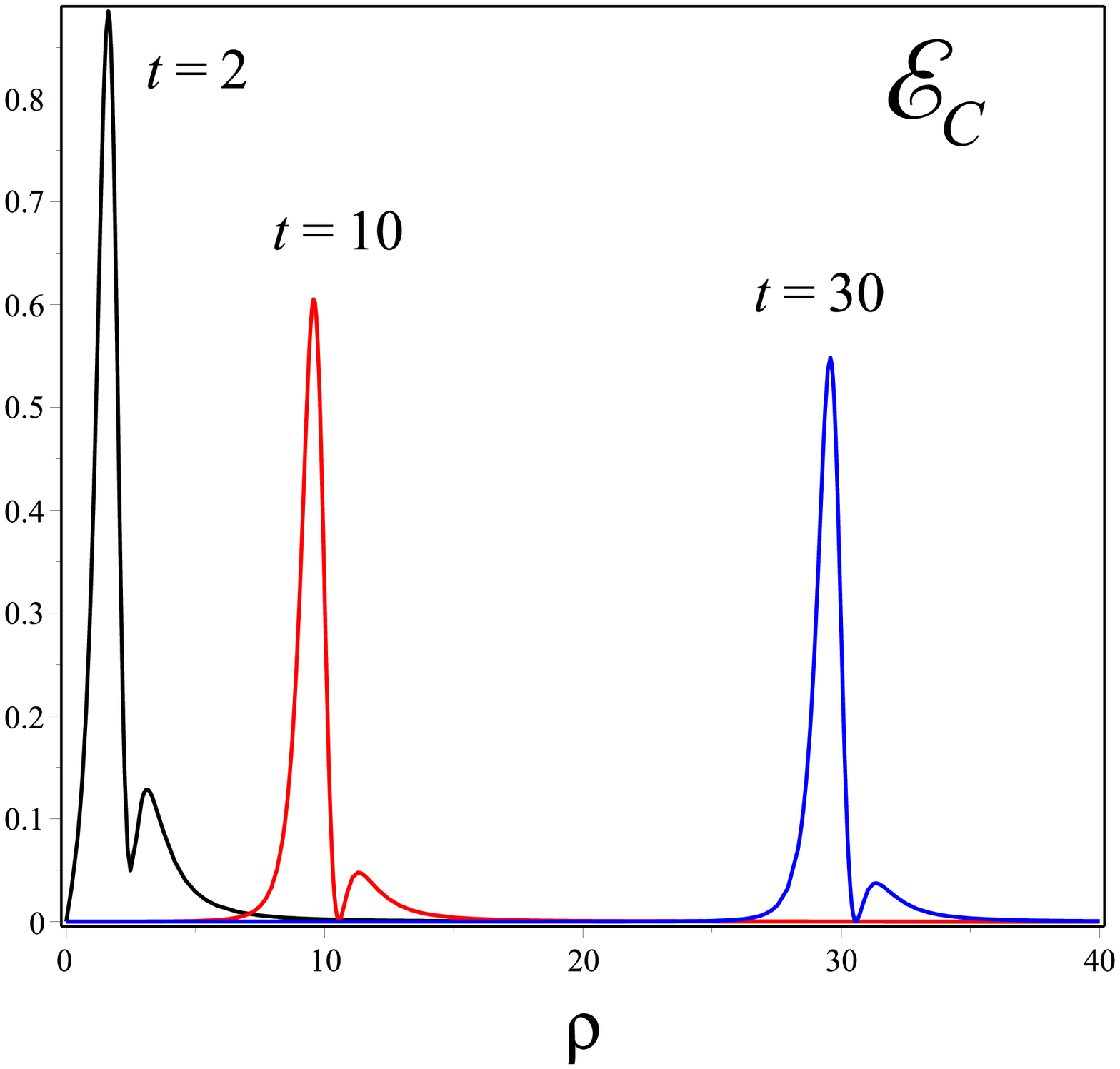}\qquad &  \includegraphics[scale=0.25]{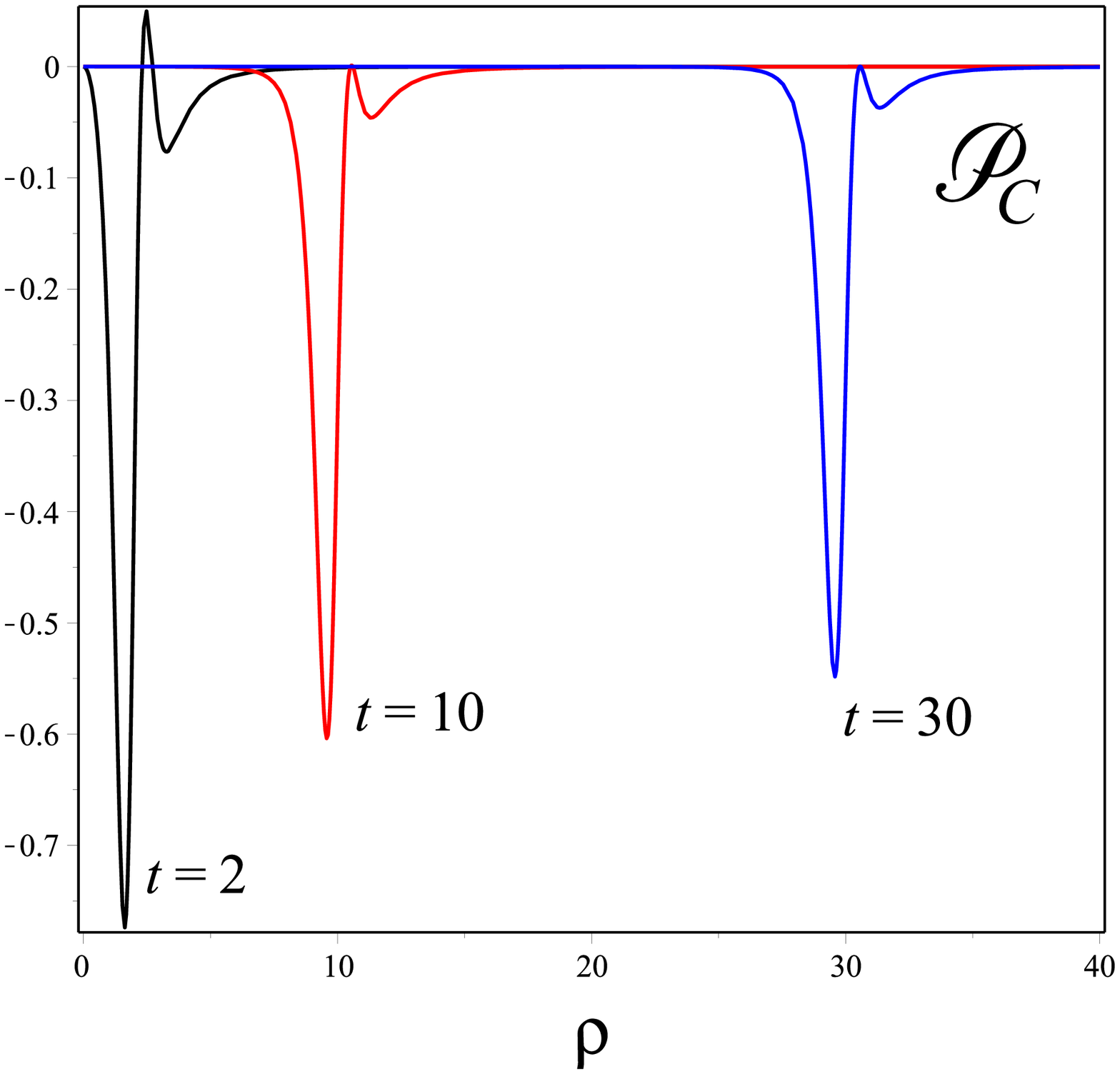}\cr
(a) & (b) \cr
\includegraphics[scale=0.25]{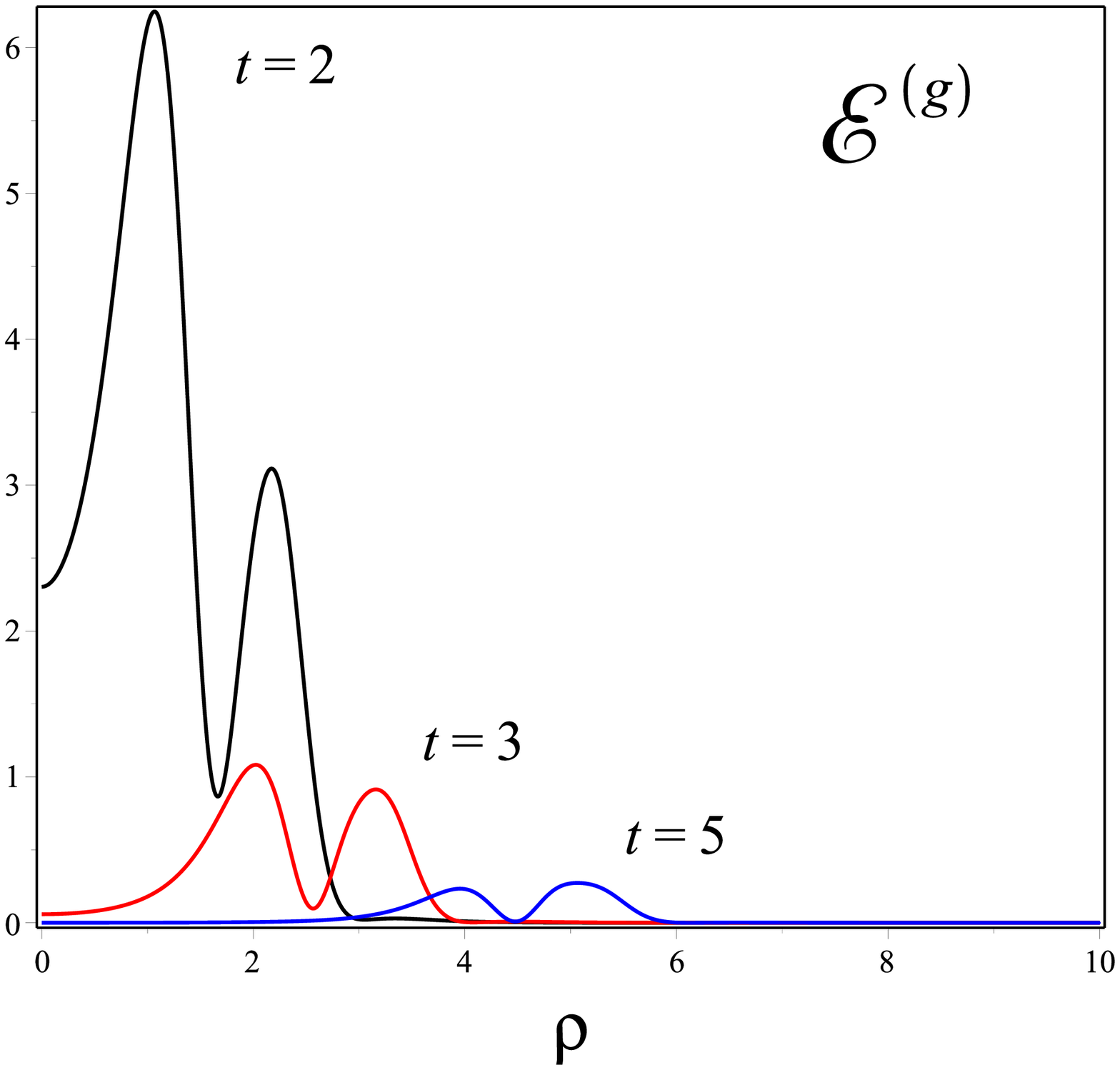}\qquad &  \includegraphics[scale=0.25]{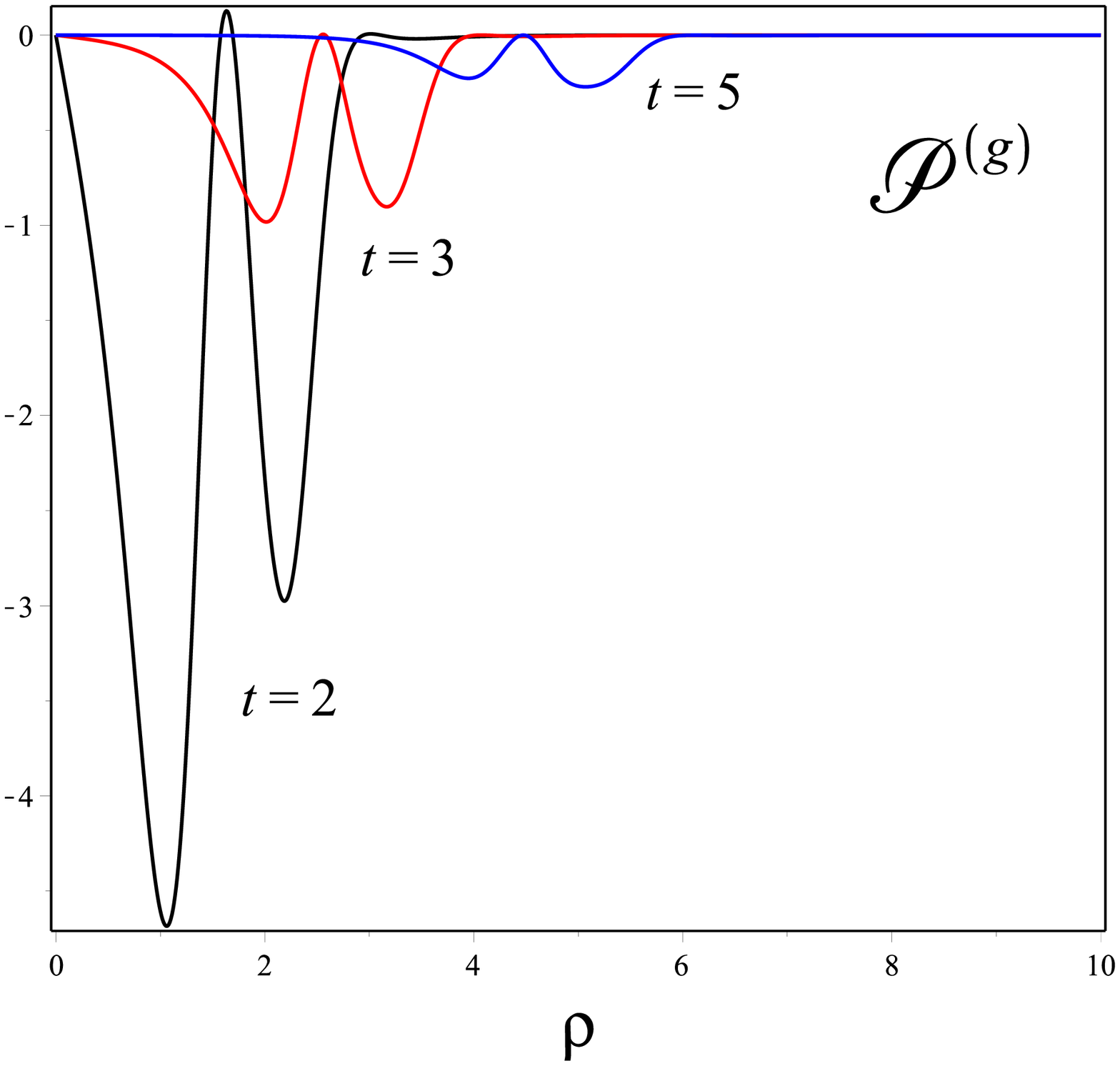}\cr
(c) & (d) 
\end{array}
\]
\caption{\label{fig:7} Weber-Wheeler-Bonnor pulse. The behavior of C-energy and C-momentum densities as functions of $\rho$ is shown in panels (a) and (b), respectively, for fixed values of $t=[2,10,30]$.
Panels (c) and (d) show instead the behavior of the super-energy and super-momentum densities for $t=[2,3,5]$.
The parameters entering the solution are chosen as $a=1=C$.}
\end{figure}

\begin{figure}
\centering
\includegraphics[scale=0.3]{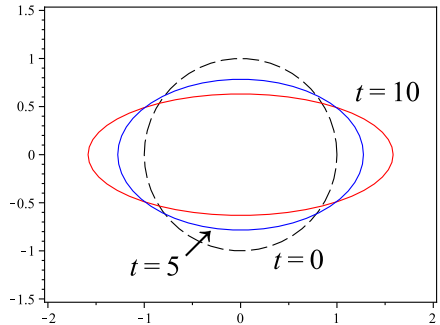}
\caption{\label{fig:8} Weber-Wheeler-Bonnor pulse. 
Deformation of a ring of particles initially at rest at $\rho=4$ induced by the pulse at different coordinate times $t=[0,5,10]$ and the same parameter choice as in Fig. \ref{fig:7}.
Further increasing the value of $t$ leaves the shape practically unchanged.
}
\end{figure}

\begin{figure}
\[
\begin{array}{cc}
\includegraphics[scale=0.25]{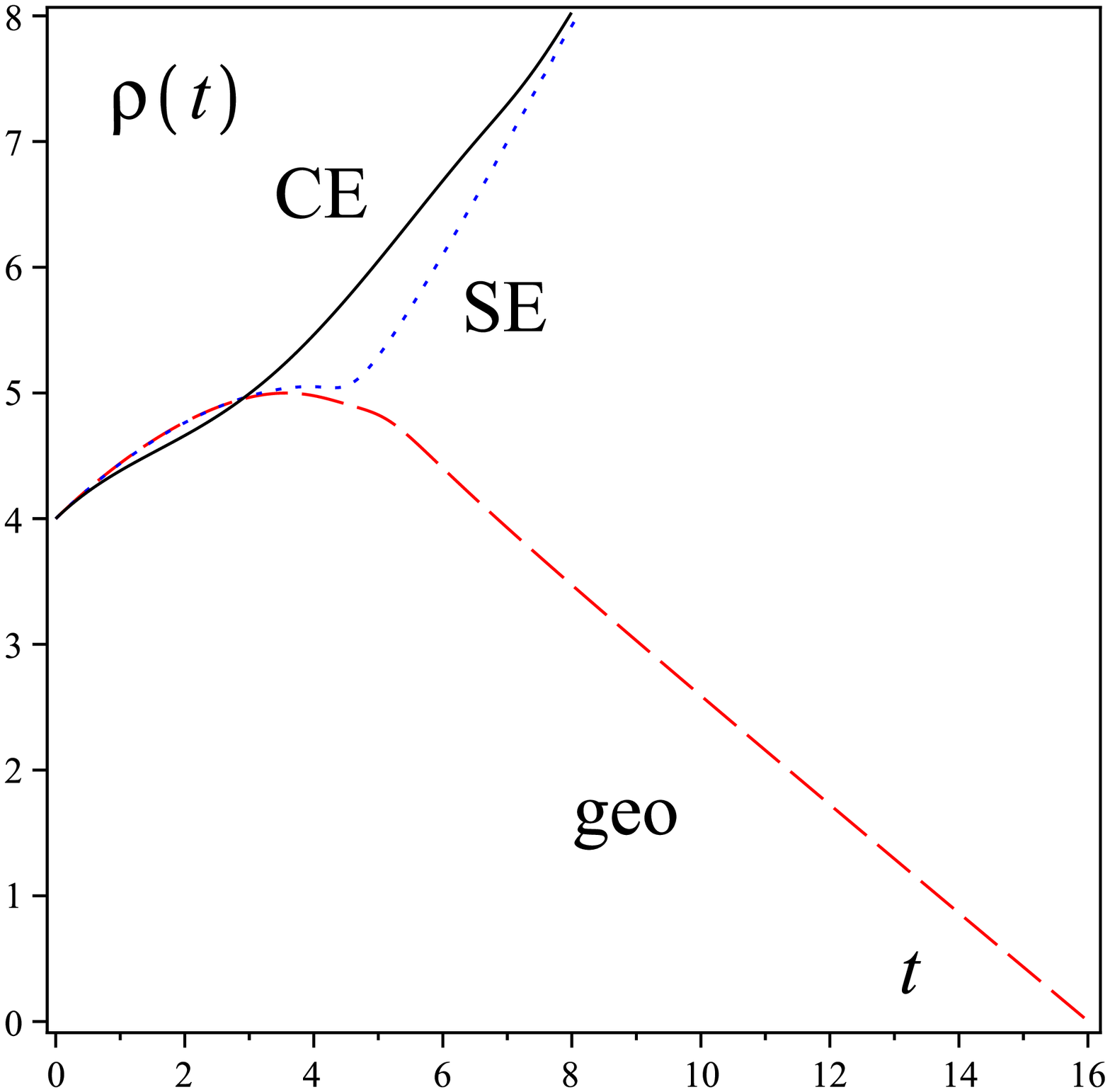}\qquad &  \includegraphics[scale=0.25]{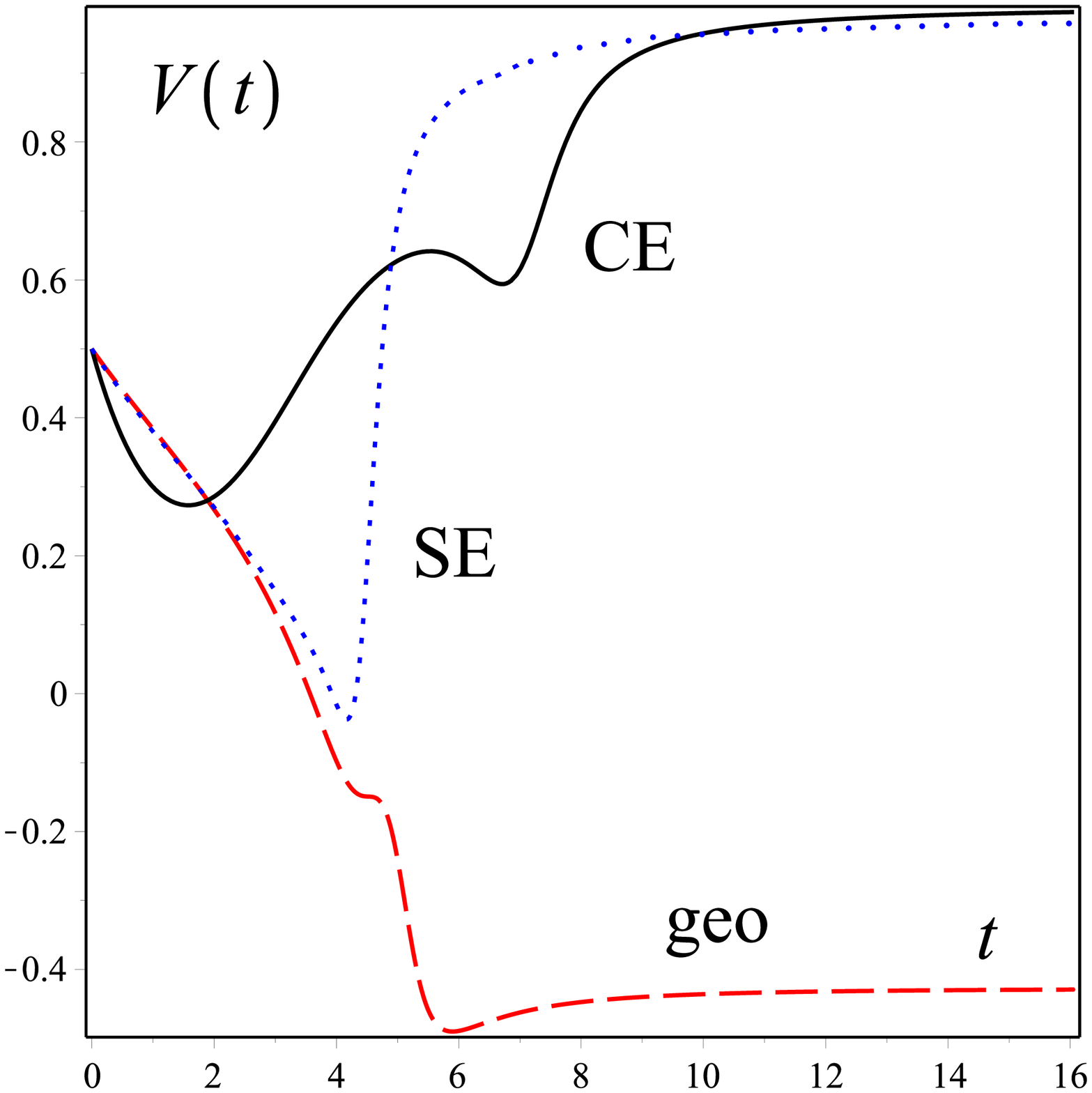}\cr
(a) & (b) 
\end{array}
\]
\caption{\label{fig:9} Weber-Wheeler-Bonnor pulse. The equations for both geodesic and accelerated radial orbits due to both C-energy and super-energy energy-momentum transfer are numerically integrated for the same parameters as in Fig. \ref{fig:7} and initial conditions $\rho(0)=4$ and $V(0)=0.5$. The additional parameter $\tilde \sigma$ is chosen as $\tilde\sigma=10$ to enhance the  external force contribution.
The radial position $\rho(t)$ (panel a) and radial velocity $V(t)$ (panel b) are plotted as functions of $t$.  
}
\end{figure}

The behavior of both C-energy and C-momentum densities as well as super-energy and super-momentum densities as functions of $\rho$ for selected values of $t$ is shown in Fig. \ref{fig:7}.
The peaks correspond to the spacetime position of the metric pulses, i.e., they are approximately located at $\rho\sim t$.
The amplitude of the peaks is generally suppressed during the evolution.
A ring of particles initially at rest at a given radius squeezes up to a certain limiting shape, as shown in Fig. \ref{fig:8}.
Finally, the numerical integration of radial equations of motion is shown in Fig. \ref{fig:9}.
Interestingly, the geodesics behave as in the previous cases, the particle reaching the axis again after a certain time (with analytical falloff of the type $\rho\sim-t$), whereas the accelerated orbits seem to escape approaching a luminal regime (with $\rho\sim t$).

% figure 7

% figure 8

% figure 9

\section{Concluding remarks}

We have explored the energy content of gravitational waves with cylindrical symmetry in terms of two different (local) notions of gravitational energy (density), well known in the literature: Thorne's C-energy and Bel-Robinson's super-energy. The first, the C-energy, is a metric-induced notion, whereas the second, the super-energy, is a curvature-induced one, and the expectation that they could bear the same physical information (our main motivation for the present analysis) is supported by the underlying wave-like equation satisfied by these solutions.
However, things are not so simple.
First of all, one should recall that both C-energy and super-energy densities are coordinate-dependent quantities with an observer-dependent character (thought having a well defined geometrical meaning), so that they transform properly when changing the coordinate system as well as when passing from one observer to another. 
Furthermore, the wave-like equations satisfied by these solutions are actually nonlinear equations, which contain all the strong-field features (i.e., the Ernst equation) of the associated gravitational field, and hence do not allow for a simple intuition of the phenomena.

We have considered the \lq\lq natural" definition of C-energy and super-energy densities in terms of static observers and adapted coordinates to the cylindrical symmetry in the case of three special solutions of the vacuum Einstein's field equations: the solution of a monochromatic cylindrical wave with two polarizations found by Chandrasekhar, the Einstein-Rosen waves and the Weber-Wheeler-Bonnor pulses.
In order to investigate the features associated with either description of local energy density of the gravitational field, we have discussed a dynamical effect on massive test particles in these backgrounds.
Indeed, it is well known that, in a first approximation beyond the geodesic motion, a particle embedded in a radiation field will interact with it by exchanging energy and momentum. This effect is called Poynting-Robertson effect, and has been largely studied in the case of a test radiation field surrounding black holes, e.g., in the presence of accretion disks.
In the same spirit, we have considered the interaction of massive particles with the gravitational field of the cylindrical waves, modeling the non-geodesic motion through a \lq\lq radiation reaction force,'' which can be naturally associated with energy and momentum flux carried by the waves.
We have built up this interaction force in a way closely related to the Poynting-Robertson prescriptions, by using either the C-energy and super-energy densities, and integrated numerically the equations for radial motion.
In both cases, as a general feature, a particle which starts moving outwards from a given position is pushed forward as a result of the interaction, eventually turning back towards the symmetry axis or even reaching an equilibrium position at a certain radius, like in the case of the Poynting-Robertson effect in black hole spacetimes. 
The results of the present analysis seem to enforce for both notions of C-energy and super-energy densities the meaning of energy density naturally associated with the waves, even if the agreement is mostly qualitative.

\appendix

\section{Jordan-Ehlers-Kundt-Kompaneets form of the most general cylindrically symmetric line element} 

The most general cylindrically symmetric line element in cylindrical-like coordinates $x^\alpha=(t,\rho,\phi,z)$ was written in Refs. \cite{Kompaneets:1958,Jordan:1960} as follows
\beq
\label{non_diag2}
ds^2 = e^{2(\gamma-\psi)}(-dt^2+d\rho^2)+e^{2\psi}\left(dz +\omega d\phi\right)^2
+\rho^2e^{-2\psi}d\phi^2\,,
\eeq
where the metric functions $\gamma$, $\psi$ and $\omega$ are related to those ($\nu$, $\chi$ and $q_2$) of Eq. \eqref{non_diag} by Eq. \eqref{newfun}, i.e.,
\beq
\nu=\gamma -\psi\,,\qquad 
\frac{\rho}{\chi}=e^{2\psi}\,, 
\qquad q_2=-\omega\,.
\eeq
The advantage of using this form of the metric is that $\psi$ and $\omega$ represent the two dynamical degrees of freedom of the gravitational field, the former corresponding to the $+$ polarization mode, the latter to 
the $\times$ mode. The function $\gamma$ plays the role of gravitational energy of the spacetime (or the C-energy, see Eq. \eqref{Cendef}). 

Vacuum Einstein's equations read
\begin{eqnarray}
\label{eqein1}
&&\psi_{tt}-\frac1{\rho}\psi_\rho-\psi_{\rho\rho}=\frac{e^{4\psi}}{2\rho^2}(\omega_t^2-\omega_\rho^2)\,,\\
\label{eqein2}
&&\omega_{tt}+\frac1{\rho}\omega_\rho-\omega_{\rho\rho}=4(\omega_\rho\psi_\rho-\omega_t\psi_t)\,,\\
\label{eqein3}
&&\gamma_\rho=\rho(\psi_t^2+\psi_\rho^2)+\frac{e^{4\psi}}{4\rho}(\omega_t^2+\omega_\rho^2)\,,\\
\label{eqein4}
&&\gamma_t=2\rho\psi_t\psi_\rho+\frac{e^{4\psi}}{2\rho}\omega_t\omega_\rho\,.
\end{eqnarray}
The orthonormal frame adapted to the family of fiducial observers with four-velocity $u=e_0=e^{-(\gamma-\psi)}\partial_t$, corresponding to the one defined in Eq. \eqref{frame}, is given by
\beq
\label{frame2}
e_1=e^{-(\gamma-\psi)}\partial_\rho\,,\quad 
e_2=\frac{e^{\psi}}{\rho}(\partial_\phi-\omega\partial_z)\,,\quad 
e_3=e^{-\psi}\partial_z\,.
\eeq

Cylindrical gravitational wave solutions are in general characterized by
the following ingoing $(I_+,I_\times)$ and outgoing $(O_+,O_\times)$ amplitudes associated with the two polarization modes $(+,\times)$ \cite{Piran:1985dk,Halilsoy:1988vz}
\begin{eqnarray}
\label{IOpmdef}
I_+&=&2(\psi_t+\psi_\rho)=2\psi_u\,,\qquad
I_\times=\frac{e^{2\psi}}{\rho}(\omega_t+\omega_\rho)=\frac{e^{2\psi}}{\rho}\omega_u
\,,\nonumber\\
O_+&=&2(\psi_t-\psi_\rho)=2\psi_v\,,\qquad
O_\times=\frac{e^{2\psi}}{\rho}(\omega_t-\omega_\rho)=\frac{e^{2\psi}}{\rho}\omega_v\,,
\end{eqnarray}
where the ingoing and outgoing null coordinates $u$ and $v$ are defined by $u=(t-\rho)/2$ and $v=(t+\rho)/2$, respectively, with $\partial_u=\partial_t+\partial_\rho$ and $\partial_v=\partial_t-\partial_\rho$.
Equations \eqref{eqein1}--\eqref{eqein4}  then 
yield then a set of four coupled first order ordinary differential equations along the characteristics $u$ and $v$ for the amplitudes
\begin{eqnarray}
\label{eqsIOpp}
I_+{}_{,u}&=&\frac1{2(v-u)}(I_+-O_+)+I_\times O_\times
=O_+{}_{,v}
\,,\nonumber\\
I_\times{}_{,u}&=&\frac1{2(v-u)}(I_\times+O_\times)-I_+ O_\times
\,,\nonumber\\
O_\times{}_{,v}&=&-\frac1{2(v-u)}(I_\times+O_\times)-O_+I_\times
\,,
\end{eqnarray}
whereas Eqs. \eqref{set2} become
\beq\fl\quad
\gamma_\rho=\frac{\rho}{8}(I_+^2+O_+^2+I_\times^2+O_\times^2)\,,\qquad
\gamma_t=\frac{\rho}{8}(I_+^2-O_+^2+I_\times^2-O_\times^2)\,.
\eeq
It is convenient to introduce the complex quantities
\beq
\label{IOcomplex}
I=I_+ + i I_\times \,,\qquad O=O_+ + i O_\times\,,
\eeq
implying, e.g.,
\beq
\gamma_\rho=\frac{\rho}{8}(|I|^2+|O|^2)\,,\qquad
\gamma_t=\frac{\rho}{8}(|I|^2-|O|^2)\,,
\eeq
where $|I|^2=I^*I $, etc. 
Moreover
\begin{eqnarray}
IO &=& (I_+O_+-I_\times O_\times)+i (I_\times O_+ +O_\times I_+)\,,\nonumber\\
I O^* &=&  (I_+O_++I_\times O_\times)+i (I_\times O_+ -O_\times I_+)\,,
\end{eqnarray}
with similar expressions for $I^*O$ and $I^* O^*$ obtained by complex conjugation of the above relations.
Equations \eqref{eqsIOpp} thus become 
\begin{eqnarray}
\label{eqsIOpp2}
I_{,u}&=&\frac1{2(v-u)}(I-O^*)-\frac{O-O^*}{2}I
\,,\nonumber\\
O_{,v}&=&\frac1{2(v-u)}(I-O^*)-\frac{I-I^*}{2}O
\,.
\end{eqnarray}

The nonvanishing frame components of the (rescaled) electric and magnetic parts of the Weyl tensor, $\tilde E(u)_{ab}=e^{2(\gamma-\psi)}E(u)_{ab}$ and $\tilde H(u)_{ab}=e^{2(\gamma-\psi)}H(u)_{ab}$, are listed below: 
\begin{eqnarray}
\tilde E(u)_{11}&=&\frac{1}{4\rho}\left[\rho \, {\rm Re}(IO^*)
+I_+-O_+\right]
\,,\nonumber\\
\tilde E(u)_{22}&=&-\frac{1}{16}\left\{
-4(I_+-O_+)_{,\rho}+\rho[I_+|I|^2-O_+|O|^2]\right.\nonumber\\
&&\left.
-3(I_+^2+O_+^2)+I_\times^2+O_\times^2+2\,  {\rm Re}(IO) 
\right\}
\,,\nonumber\\
\tilde E(u)_{23}&=&\frac{1}{16}\left\{
-4(I_\times-O_\times)_{,\rho}+\rho[I_\times |I|^2  -O_\times  |O|^2 ]\right.\nonumber\\
&&\left.
+2  {\rm Im}(IO) 
-4(I_+I_\times +O_+O_\times)
\right\}
\,,
\end{eqnarray}
with $E(u)_{33}=-E(u)_{11}-E(u)_{22}$, and
\begin{eqnarray}
\tilde H(u)_{11}&=&\frac{1}{4\rho}\left[\rho \, {\rm Im}(IO^*)
+I_\times+O_\times\right]
\,,\nonumber\\
\tilde H(u)_{22}&=&-\frac{1}{16}\left\{
-4(I_\times-O_\times)_{,\rho}+8O_\times{}_{,t}
+\rho[I_\times |I|^2+O_\times |O|^2]\right.\nonumber\\
&&\left.
+8O_+I_\times-4(I_+I_\times-O_+O_\times)+\frac{4}{\rho}(I_\times +O_\times)
\right\}
\,,\nonumber\\
\tilde H(u)_{23}&=&\frac{1}{16}\left\{
4(I_+-O_+)_{,\rho}-8O_+{}_{,t}
-\rho[I_+|I|^2+O_+|O|^2]\right. \nonumber\\
&&\left.
+8I_\times O_\times-I_\times^2+O_\times^2+3(I_+^2-O_+^2)+\frac{4}{\rho}(I_+-O_+)
\right\}
\,,\nonumber\\
\end{eqnarray}
with $H(u)_{33}=-H(u)_{11}-H(u)_{22}$.
Introducing then the complex combination
\beq
\tilde Z(u)_{ab}=\tilde E(u)_{ab}+ i \tilde H(u)_{ab}
\eeq
simplifies only some expressions, e.g.,
\beq
\tilde Z(u)_{11}=\frac{1}{4\rho} \left[\rho\, (IO^*)+I-O^*  \right]\,.
\eeq

\end{document}